\documentclass[%
 reprint,
 amsmath,
 amssymb,
 aps,
 prl,
floatfix,
superscriptaddress
]{revtex4-2}

\usepackage{graphicx}
\usepackage{dcolumn}
\usepackage{bm}
\usepackage[colorlinks=true, allcolors=blue]{hyperref}
\usepackage{color}
\usepackage{comment}
\usepackage{float}
\usepackage{subcaption}
\usepackage{amsmath}
\usepackage{amssymb}
\usepackage[mathlines]{lineno}

\usepackage{pgffor}

\usepackage{bibunits}
\defaultbibliographystyle{apsrev4-1}
\usepackage{etoolbox}

\makeatletter
\newcommand*{\newbibstartnumber}[1]{%
  \apptocmd{\thebibliography}{%
    \global\c@NAT@ctr #1\relax
    \addtocounter{NAT@ctr}{-1}%
  }{}{}%
}
\makeatother

\begin{document}


\title{
Electron-only magnetic reconnection and inverse magnetic-energy transfer at sub-ion scales
}

\author{Zhuo Liu}
\affiliation{%
Plasma Science and Fusion Center, Massachusetts Institute of Technology, Cambridge, MA 02139, USA
}%
\author{Caio Silva}%
\affiliation{%
Plasma Science and Fusion Center, Massachusetts Institute of Technology, Cambridge, MA 02139, USA
}%
\author{Lucio M. Milanese}
\affiliation{%
Proxima Fusion GmbH, Munich, 81671, Germany}%
\author{Muni Zhou}
\affiliation{School of Natural Sciences$,$ Institute for Advanced Study$, $ Princeton$,$ NJ 08544$,$ USA}
\author{Noah R. Mandell}
\affiliation{%
Princeton Plasma Physics Laboratory, Princeton, NJ 08543, USA
}%
\author{Nuno F. Loureiro}
\affiliation{%
Plasma Science and Fusion Center, Massachusetts Institute of Technology, Cambridge, MA 02139, USA
}%

\date{\today}%

\begin{abstract}
We derive, and validate numerically, an analytical model for electron-only magnetic reconnection applicable to strongly magnetized plasmas. Our model predicts sub-ion-scale reconnection rates significantly higher than those pertaining to large-scale reconnection, aligning with recent observations and simulations. 
We apply this reconnection model to the problem of inverse magnetic energy transfer at sub-ion scales. 
We derive time-dependent scaling laws for the magnetic energy decay and the typical magnetic structure dimensions that differ from those previously found in the MHD regime. 
These scaling laws are validated via two- and three-dimensional simulations, demonstrating that sub-ion scale magnetic fields can reach large, system-size scales via successive coalescence. 
\end{abstract}
\maketitle


\paragraph{Introduction.} 
Decaying turbulence in plasmas is sometimes accompanied by the transfer of energy from small to large scales; a so-called inverse transfer. 
Though poorly understood, this phenomenon is thought to be crucial to magnetic field evolution at cosmological scales~\cite{banerjee2004evolution,durrer2013cosmological,hosking2023cosmic} as well as in various astrophysical systems such as the solar wind~\cite{chen2011anisotropy} and the heliosphere~\cite{khabarova2015small,drake2012power}.
The standard explanation for inverse energy transfer appeals to the conservation of net magnetic helicity~\cite{pouquet1978two,matthaeus1986turbulent,christensson2001inverse}. However, recent studies have shown that this process can also occur in non-helical turbulent systems~\cite{brandenburg2015nonhelical, reppin2017nonhelical}. 
It has been proposed that this may be due to the successive merger of magnetic structures via magnetic reconnection, thus transferring energy to progressively larger scales ~\citep{zhou2019magnetic,zhou2020multi,bhat2021inverse,zhou2021statistical,hosking2021reconnection,milanese2023}.  

So far, investigations of this topic have mostly been restricted to the resistive MHD framework.
This implies they cannot be directly applied to 
the (mostly) collisionless environments often encountered in astrophysical contexts.
For example, recent \textit{in-situ} spacecraft measurements have provided substantial evidence for the existence of numerous rope-like magnetic structures at ion scales, with substructures within these flux ropes observed at electron scales (e.g.,~\cite{stawarz2018intense,zhong2020direct,huang2021electron,hasegawa2023ion}).
As such, the existing understanding of inverse energy cascade needs to be extended to account for (i) kinetic modifications to turbulent dynamics; and (ii) changes to magnetic reconnection --- as compared to resistive MHD systems.

The latter is of particular importance given the observation that  the physics of reconnection between sub-ion-scale structures is expected to be substantially different:  
when the reconnection region is sufficiently small, the ions can become unresponsive, and a transition to so-called ``electron-only reconnection'' is expected.
This variant of magnetic reconnection has been recently observed in the Earth's magnetosphere~\citep{phan2018electron,stawarz2019properties}, and is numerically found to exhibit substantially higher reconnection rates compared to conventional scenarios~\citep{jain2012electron,sharma2019transition,califano2020electron,arro2020statistical,vega2020electron,greess2022kinetic,guan2023reconnection}.

The goal of this Letter is thus to generalize the current resistive-MHD-based understanding of inverse energy transfer via magnetic reconnection to kinetic, collisionless plasma environments.

\paragraph{Equations.} 
We employ a two-field isothermal fluid model valid for low-beta plasmas~\citep[e.g.,][]{schekochihin2009astrophysical,zocco2011reduced,passot2018gyrofluid, meyrand2021violation, zhou2023spectrum}.
The dynamics of this system are described by equations for the perturbed electron density $\delta n_e$ and the parallel (to the guide-field $B_z$) vector potential $A_z$, which read~\footnote{These equations are the isothermal limit of the Kinetic Reduced Electron Heating Model (KREHM), a rigorous analytical framework derived from gyrokinetics under the condition of low electron plasma beta ($\beta_e \sim m_e/m_i$, where $\beta_e$ is the ratio of the electron to the magnetic pressure, and $m_e$ and $m_i$ are the electron and proton masses)~\citep{zocco2011reduced}.
See Supplemental Material for more details and results in the non-isothermal limit.}:
\begin{align}
\frac{1}{n_{0e}} \frac{d\, \delta n_e}{d t} = \hat{\mathbf{b}}\cdot \nabla \frac{e}{c m_e} d_e^2 \nabla_{\perp}^2 A_z, \label{eq:ne} \\
\frac{d}{d t}\left(A_z-d_e^2 \nabla_{\perp}^2 A_z\right)= - c\frac{\partial \varphi}{\partial z} + \hat{\mathbf{b}}\cdot \nabla \frac{\delta n_e}{n_{0 e}}.
\label{eq:Az}
\end{align}
Here, $\hat{\mathbf{b}}\cdot \nabla \equiv \partial/\partial_z -1/B_z \{A_z,...\}$ is the parallel (to the total field) gradient operator, and $d/dt \equiv \partial/\partial t+c/B_z\{\varphi,...\}$ denotes the convective time derivative, with $\varphi$ the electrostatic potential; both operators feature a Poisson bracket, defined as $\{P,Q\} \equiv \partial_x P \partial_y Q-\partial_y P \partial_x Q$.
These equations are closed via 
the gyrokinetic Poisson law~\citep{krommes2002fundamental},
\begin{equation}
    \label{eq:gk_poisson}
    \frac{\delta n_e}{n_{0e}}=\frac{T_{0e}}{T_{0i}}(\hat{\Gamma}_0-1)\frac{e\varphi}{T_{0e}},
\end{equation}
where $\hat{\Gamma}_0$ denotes the inverse Fourier transform of $\Gamma_0(\alpha)=I_0(\alpha)e^{-\alpha}$, with $I_0$ the zeroth-order modified Bessel function of the first kind and $\alpha = k_\perp^2 \rho_i^2/2$, where $\rho_i = v_{{\rm th}i}/\Omega_i$ is the ion Larmor radius, with $v_{{\rm th}i}=\sqrt{2T_{0i}/m_i}$ the ion thermal velocity and $\Omega_i=eB_z/m_ic$ the ion Larmor frequency.
Lastly, $d_e=c/\omega_{pe}$ is the electron skin depth, with $\omega_{pe}=\sqrt{4\pi n_{0e}e^2/m_e}$ the electron plasma frequency.
In the following, we refer to ``perpendicular'' as the direction perpendicular to the guide field $B_z$, i.e., the $xy$ plane.

\paragraph{Electron-only Reconnection.}~We now use these equations to derive a two-dimensional reconnection model valid at scales below the ion Larmor radius.
We consider two identical magnetic islands with typical magnetic field $B_\perp$ and radius $R$ such that $d_e < R < \min(\rho_i,\rho_s)$~\footnote[38]{We use $<$ instead of $\ll$ in order not to interfere with the ordering under which Eqs.~(\ref{eq:ne}-\ref{eq:gk_poisson}) are derived (namely, $\beta \sim m_e/m_i$ so that $\rho_i \sim d_e$, see~\citep{zocco2011reduced}). 
We note that $\beta$ does not have to be numerically close to $m_e/m_i$; all that is formally required is for $\rho_i/d_e$ not to be so large as to interfere with the asymptotic expansion parameters. Empirically, it is known that Eqs.~(\ref{eq:ne}-\ref{eq:gk_poisson}) remain valid at least up to $\beta \sim 0.1$ (see ~\citep{grovselj2017fully}). In practice, a maximum ratio of $\rho_i/d_e\approx 20$ is acceptable if adopting the real proton-electron mass ratio.}, merging in the $x$ direction, where $\rho_s = \rho_i/\sqrt{2T_{0i}/T_{0e}}$ is the ion sound Larmor radius.
The total magnetic flux to be reconnected is $A_{z,\rm{total}} = B_{\perp} R$, and the time for this process to occur (reconnection time) is denoted as $\tau_{\rm rec}$.
The width of the current sheet between the islands during reconnection is denoted as $\delta_x$, assumed comparable to $d_e$ (the scale at which the frozen flux condition is broken in our model, and the realizability is confirmed via direct numerical simulations below),
and the length of the current sheet is denoted as $\delta_y$.
In the current sheet, $A_z = A_{z,\rm{sheet}} \sim B_{\perp} \delta_x$, 
consistent with an upstream reconnecting field $B_y = \partial_x A_z \sim B_\perp$. 
The out-of-plane current in the sheet is then $-(4\pi/c) J_z =\nabla_{\perp}^2 A_z \sim B_{\perp}/\delta_{x}$.

In the sub-ion regime ($k_{\perp}\rho_i >1$), ions are demagnetized, and   Eq.~\eqref{eq:gk_poisson} reduces to $\delta n_e/n_{0e} \approx -e\varphi/T_{0i}$; that is, electrons follow a Boltzmann response~\citep{zocco2011reduced}.
Thus, the advection of $\delta n_e$ term in Eq.~\eqref{eq:ne} vanishes.   
The advection of $A_z$ term in Eq.~\eqref{eq:Az}, i.e., $\{\varphi,A_z\}$, becomes $- (cT_{0i}/eB_z) \{\delta n_e/n_{0e}, A_z\}$, which can be combined with the R.H.S. of Eq.~\eqref{eq:Az}.
Taking $\partial_x \sim 1/\delta_x$ and $\partial_y \sim 1/\delta_y$ and balancing both sides of Eq.~\eqref{eq:ne} within the sheet, 
we find $(\delta n_e/n_{0e})/\tau \sim (e d_e^2/B_zcm_e) (B_\perp^2/\delta_x \delta_y)$,
where $\tau$ is the time it takes to reconnect the magnetic flux $A_{z, \rm{sheet}}$.
A similar term-balancing procedure for  Eq.~\eqref{eq:Az} yields 
$A_z / \tau \sim (c(T_{0e}+T_{0i})/eB_zn_{e0}) (B_{\perp}\delta n_e/\delta_y)$.
Note that the electron inertia term does not affect this balance since $d_e^2 \nabla_{\perp}^2A_{z} \sim (d_e/\delta_x)^2  A_{z} \sim A_{z}$.

Combining these results, we find $\tau \sim (1/\sqrt{1+T_{0i}/T_{0e}}) (\delta_y/\rho_s) (d_e/v_{A,\perp})$, where $v_{A,\perp} = B_{\perp}/\sqrt{4\pi n_0 m_i}$ is the (ion) Alfv\'en speed.
The last step is to relate the reconnection time, $\tau_{\rm rec}$, and $\tau$ via $\tau_{\rm rec} / \tau \sim A_{z, \rm{total}}/  A_{z, \rm{sheet}}\sim R/d_e$.
Writing the reconnection time in the form of $\tau_{\rm rec} \sim \mathcal{R}_{\rm rec}^{-1} R/v_{A,\perp}$, we obtain the normalized reconnection rate: 
\begin{equation}
    \label{eq:rec_rate}
    \mathcal{R}_{\rm rec} \sim \sqrt{1+\frac{T_{0i}}{T_{0e}}}\frac{\rho_s}{\delta_y} \sim  \sqrt{1+\frac{T_{0i}}{T_{0e}}}\frac{\rho_s}{R},
\end{equation}
where we have assumed $\delta_y\sim R$ (see Supplemental Material for validation).
When $\rho_s > R$, the (normalized) reconnection rate can become much larger than the value of $\sim 0.1$ typically associated with collisionless reconnection~\cite{birn2001GEM,cassak2017review,liu2017does} (but may not exceed $\sim \rho_s/d_e$).
See Appendix A for further details.

The outflow velocity, $u_{\rm out}$, can be estimated as the $E \times B$ drift velocity, i.e., $u_{\rm out} \sim (c/B_x)E_z \sim cR/(d_eB_{\perp}) (B_{\perp}R/c \tau_{\rm rec}) \sim \sqrt{1+T_{0i}/T_{0e}}(\rho_s/d_e)v_{A,\perp} \sim \sqrt{1+T_{0i}/T_{0e}} \sqrt{\beta_e}v_{Ae,\perp}$, with $v_{Ae,\perp}\equiv B_{\perp}/\sqrt{4\pi n_{0e}m_e}$ the electron Alfv\'en speed based on perpendicular magnetic field.
The conclusion that the outflow velocity is proportional to electron Alfv\'en speed is consistent with previous laboratory observations~\citep{shi2022laboratory}.
Therefore, if instead we normalize the reconnection time to the outflow time, i.e., $\tau_{\rm rec} \sim \mathcal{R}_{\rm rec,e}^{-1}R/u_{\rm out}$, we obtain
\begin{equation}
    \mathcal{R}_{\rm rec,e} \sim {d_e}/{R};
\end{equation}
i.e., the aspect ratio of the reconnecting current sheet.

We now proceed to validate this model via direct numerical simulations. 
Eqs.~(\ref{eq:ne},\ref{eq:Az},\ref{eq:gk_poisson}) are solved numerically using the pseudo-spectral code GX~\cite{mandell2018laguerre,mandell2022gx}.
Hyper-diffusive terms of the form $\nu_H \nabla_{\perp}^6$ are added to the R.H.S. of Eqs.~(\ref{eq:ne},\ref{eq:Az}), with $\nu_H = 1/\Delta t (\Delta x/\pi)^6$; this value ensures that these terms are only significant at scales comparable to the grid scale. These terms are required to prevent the nonlinear unbounded thinning of the current sheet~\citep{rogers1996collisionless,grasso2000ion}.
These details apply to all the simulations we include in this Letter.

In what follows, all quantities are presented in dimensionless units.
The simulations are performed in a doubly periodic box of dimensions $L_x \times L_y$, where $L_x = L_y =2\pi$.
The initial condition consists of two identical magnetic islands of Gaussian shape, specified as  $A_z = A_0 \exp(- (2\pi a (x - 0.25L_x)/L_x)^2 - (2\pi a (y - 0.5L_y)/L_y)^2)+ A_0 \exp(- (2\pi a (x - 0.75L_x)/L_x)^2- (2\pi a (y - 0.5L_y)/L_y)^2) $ ~\footnote{This initial configuration is approximately an equilibrium but it is unstable to coalescence instability~\cite{finn1976coalescence}. We choose this configuration instead of the typical Harris sheet~\cite{harris1962plasma} for consistency with the study of the inverse magnetic energy transfer problem discussed below. However, we do not expect the results to be sensitive to the specific configuration chosen.}.
We set $a=1.5$ so that the root-mean-squared width of each island is about $1.1$, i.e., $R\approx 0.55$.
Multiple simulations are performed by varying the values of $\rho_i$ (we set $T_{i0}/T_{e0}=1$ in all the simulations so $\rho_s = \rho_i/\sqrt{2}$) and $d_e$ to encompass both the regime where $R > \rho_i> d_e$ (ion-coupled reconnection) and $\rho_i \gtrsim R > d_e$ (electron-only reconnection). 
All simulations have $\rho_i/d_e \leq 20$ so as to conform to the low-beta ordering that underlies the equations we use.
The numerical resolution is adjusted across simulations to resolve the $d_e$ scale adequately (see Supplemental Material for more information on simulation parameters).

\begin{figure}[!htp]
\includegraphics[width=0.45\textwidth]{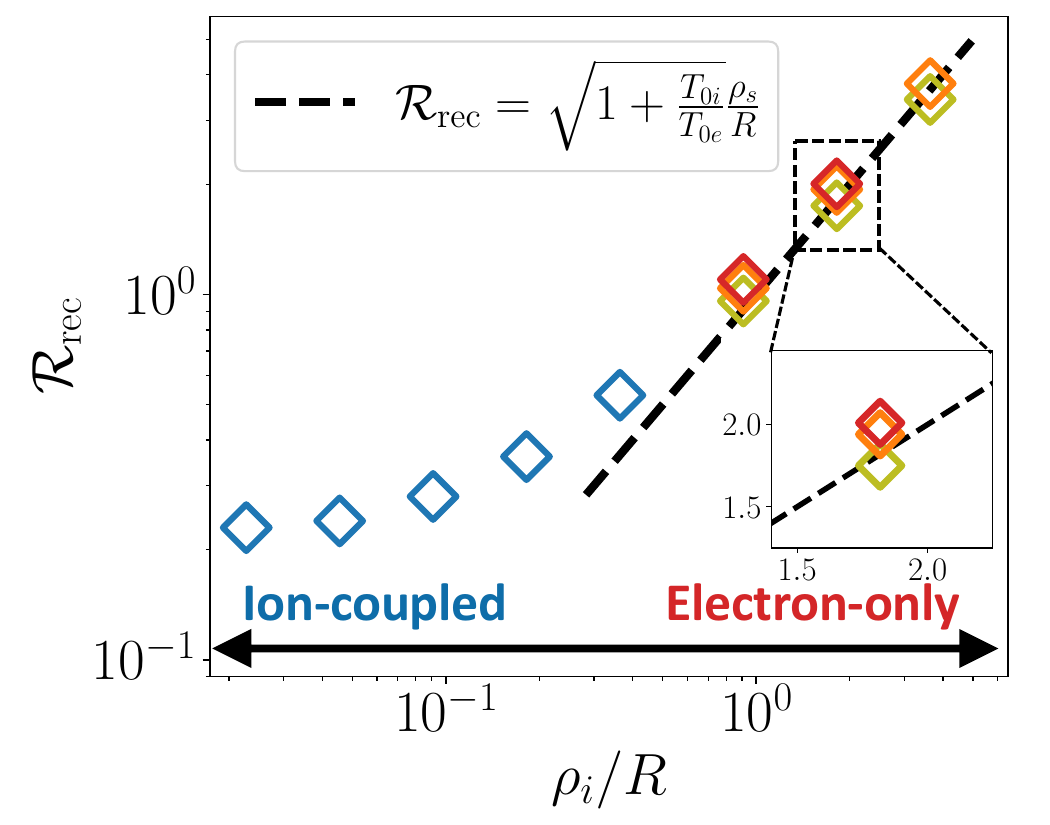}
\caption{\label{fig:1.rec_rate} Dimensionless peak reconnection rates, $\mathcal{R}_{\rm rec}$, from simulations with different values of $\rho_i/R$. 
In the electron-only regime ($\rho_i \gtrsim R > d_e$),  red, orange, and yellow symbols refer to simulations with $d_e/R=0.09,018,0.36$, respectively. In the ion-coupled regime ($R > \rho_i> d_e$), we set $d_e/\rho_i=0.1$ (except for the leftmost data-point, where $d_e/\rho_i = 0.2$). The dashed line is Eq.~(\ref{eq:rec_rate}) (with $T_{i0}/T_{e0}=1$). }
\end{figure}

For each simulation, we calculate the reconnection rate as the time derivative of the magnetic flux between the X- and the  O-points.
The peak reconnection rates are plotted in Fig.~\ref{fig:1.rec_rate} against the parameter $\rho_i/ R$.
Along the horizontal axis, the progression from left to right signifies a transition from MHD to sub-$\rho_i$ scales in terms of the initial island size.
At MHD scales, we observe that the reconnection rate converges to a constant value $\sim 0.2$, somewhat higher than the often quoted value of $0.1$~\cite[e.g.,][]{birn2001GEM,Isobe2005MeasurementOT,Zenitani2008,Drake2008,Bessho_2012,cassak2017review,Chen2017,liu2017does} but consistent with previous results for collisionless reconnection in the strong guide-field limit~\cite{loureiro2013fast,numata2015ion}. 
However, upon transition to sub-ion scales (i.e., $R \lesssim \rho_i$), the reconnection rates exhibit a substantial increase, aligning almost perfectly with our model's prediction, Eq.~(\ref{eq:rec_rate}), represented by the dashed line in Fig.~\ref{fig:1.rec_rate}. 
Moreover, we notice that the agreement with Eq.~(\ref{eq:rec_rate}) is already very good when $R\approx \rho_i$ and $R/d_e \lesssim 5$. Therefore, large scale separation between $\rho_i$ and $d_e$ does not appear to be necessary for Eq.~(\ref{eq:rec_rate}) to be valid (see~[38]).
Consistently, at fixed $\rho_i/R$, the reconnection rate depends only weakly on the value of $d_e$ (and this dependence weakens as $d_e$ becomes smaller); while we confirm that  $\delta_x \sim d_e$ approximately holds for each simulation.

We also note that the reconnection rate starts to significantly deviate from the large-scale (ion-coupled) value ($\sim 0.2$) at around $R \sim 10\rho_i$.
This observation is consistent with the analysis in~\cite{mallet2020onset} and with simulation results~\cite{sharma2019transition}, where it is predicted that ions start to decouple from the reconnecting magnetic field at scales of around 10 ion-scales.
However, it is not until $R \sim \rho_i$ that the reconnection rate perfectly aligns with Eq.~\eqref{eq:rec_rate}.

Lastly, we note that it is possible to extend our analytical model to the $\beta\sim 1$ case --- see Supplemental Material, Section 2. In that regime, we again find that the reconnection rate is proportional to $1/R$.

\paragraph{Magnetic Flux Tube Coalescence.} We now apply the electron-only reconnection model to the problem of inverse transfer of magnetic energy at sub-ion scales within the framework of Eqs.~(\ref{eq:ne}-\ref{eq:gk_poisson}).
We first address two-dimensional dynamics in the perpendicular plane.
Consider two identical circular magnetic islands of radius $R$ such that $\rho_i \gtrsim R > d_e$. 
Each carries the magnetic flux $R B_{\perp}$, which is assumed to be conserved by their merger~\cite{zhou2019magnetic,zhou2020multi}.  
The lifetime of these two islands is determined by the reconnection time, $\tau_{{\rm rec}} \sim (R/\rho_s) (R/v_{A,\perp})$, as shown above.
Using the conservation of magnetic flux (and noting that $v_{A,\perp} \propto B_{\perp}$ and $\rho_s$ is independent of $B_{\perp}$ under the strong guide field condition), we find $R \propto B_{\perp}^{-1}$ and, thus, $\tau_{{\rm rec}} \propto B_{\perp}^{-3}$.

Now considering a ``sea'' of (volume-filling) magnetic islands with similar sizes in the sub-$\rho_i$ regime and assuming the merging process to occur sequentially between island pairs, the evolution of total magnetic energy can be described by
\begin{equation}
\frac{d \mathcal{E}}{d t} \propto \frac{d B_{\perp}^2}{d t } \sim -\frac{B_{\perp}^2}{\tau_{\rm rec}} \propto - B_{\perp}^5 \propto \mathcal{E}^{5/2}.
\label{eq:evolution}
\end{equation}
We thus find $\mathcal{E} = [a(t - t_0) + \mathcal{E}_0^{-3/2}]^{-2/3}$, where $t_0$ is the lifetime of the first island generation, $a$ is a coefficient, and $\mathcal{E}_0$ is the initial magnetic energy.
The perpendicular coherence scale and magnetic field strength can be derived in a similar way. The resulting long-term ($\tilde{t} \equiv t/t_0 \gg 1$) evolution scaling laws are
\begin{equation}
k_{\perp} \propto \tilde{t}^{-1 / 3}, \quad B_\perp \propto \tilde{t}^{-1 / 3}, \quad \mathcal{E} \propto \tilde{t}^{-2/3}.
\label{eq:scale_subrhoi}
\end{equation}

When extending to the three-dimensional case (with a strong guide field), a magnetic island of radius $R$ and magnetic field $B_{\perp}$ becomes a magnetic flux tube with radius $R$, poloidal magnetic field $B_{\perp}$, and length $l$ in the direction of the local mean field.
The arguments above leading to Eq.~(\ref{eq:scale_subrhoi}) should remain valid (which we will confirm via direct numerical simulation). 
The length of the flux tube $l$ can be estimated from the assumption that the inverse linear time scale, $\gamma_{l} \sim \omega_{KAW} \sim k_{\perp}\rho_s k_{\|}v_A$ (with $k_{\perp} \sim 1/R$ and $l\sim 1/k_\|$), should be comparable to the inverse nonlinear time scale, $\gamma_{nl}^{-1}$ (critical balance~\cite{goldstein1995properties}), where
$\gamma_{nl}^{-1}$ should be the smallest of the reconnection time, $\tau_{\rm rec}$, and the perpendicular advection time, $R/u_{\perp}$, with $u_{\perp}\sim (c/B_z)\varphi/R$ being the in-plane ($\mathbf{E} \times \mathbf{B}$) velocity that pertains to Eqs.~(\ref{eq:ne}-\ref{eq:Az}).
By assuming scale-by-scale equipartition between the density and magnetic energy fluctuations, $(\delta n_e/n_{0e})^2n_{0e}T_{0e} \sim |\nabla_{\perp}A_z|^2/8\pi$~\citep{zhou2023spectrum} (see Section 5 of the Supplemental Material for validation of this hypothesis), together with Eq.~\eqref{eq:gk_poisson} in the limit $k_{\perp}\rho_i > 1$, we find $\varphi \sim \sqrt{T_{0i}/T_{0e}}(\rho_i v_A/c) B_{\perp}$ and, thus, $R/u_{\perp} \sim (R/\rho_s) (R/v_{A,\perp}) \sim \tau_{\rm rec}$.
Therefore, we conclude that the nonlinear advection time and reconnection time are comparable within this regime, i.e., $\gamma_{nl} \sim \tau_{\rm rec}^{-1} \sim k_{\perp}\rho_s k_{\perp}v_{A,\perp}$.
This conclusion differs from that in the MHD regime, where reconnection is always slower~\cite{zhou2020multi}.
Finally, balancing $\gamma_{l}$ with $\gamma_{nl}$ yields
\begin{equation}
    l \sim { B_z}/({k_{\perp}B_{\perp}}) \propto \tilde{t}^{2/3}.
    \label{eq:length}
\end{equation}

To verify these predictions, we conducted a three-dimensional simulation in a periodic domain with dimensions $L_x \times L_y \times L_z$, where $L_x = L_y = 2\pi$ and $L_z = 4L_x$ (in normalized units).
We specify the initial equilibrium as $A_z(x,y,z,t = 0) = A_{z0} \cos(k_0 x)\cos(k_0 y)$ with $k_0 = 10$ with $A_{z0}=0.1$,  yielding a $2k_0 \times 
2k_0$ static array of magnetic flux tubes with sequentially opposite polarities.
We set $\rho_i = 1.0$ (and $T_{i0} = T_{e0}$) and $d_e= 0.04$.
These choices ensure that all flux tubes satisfy $d_e < R < \rho_i$, and thus Eqs.~(\ref{eq:scale_subrhoi}-\ref{eq:length}) should be valid after an initial transient.
We note that because all initial magnetic structures are small-scale,  magnetic energy can only reach larger scales through inverse transfer, that is, magnetic island mergers.

\begin{figure}[!htbp]
\includegraphics[width=0.48\textwidth]{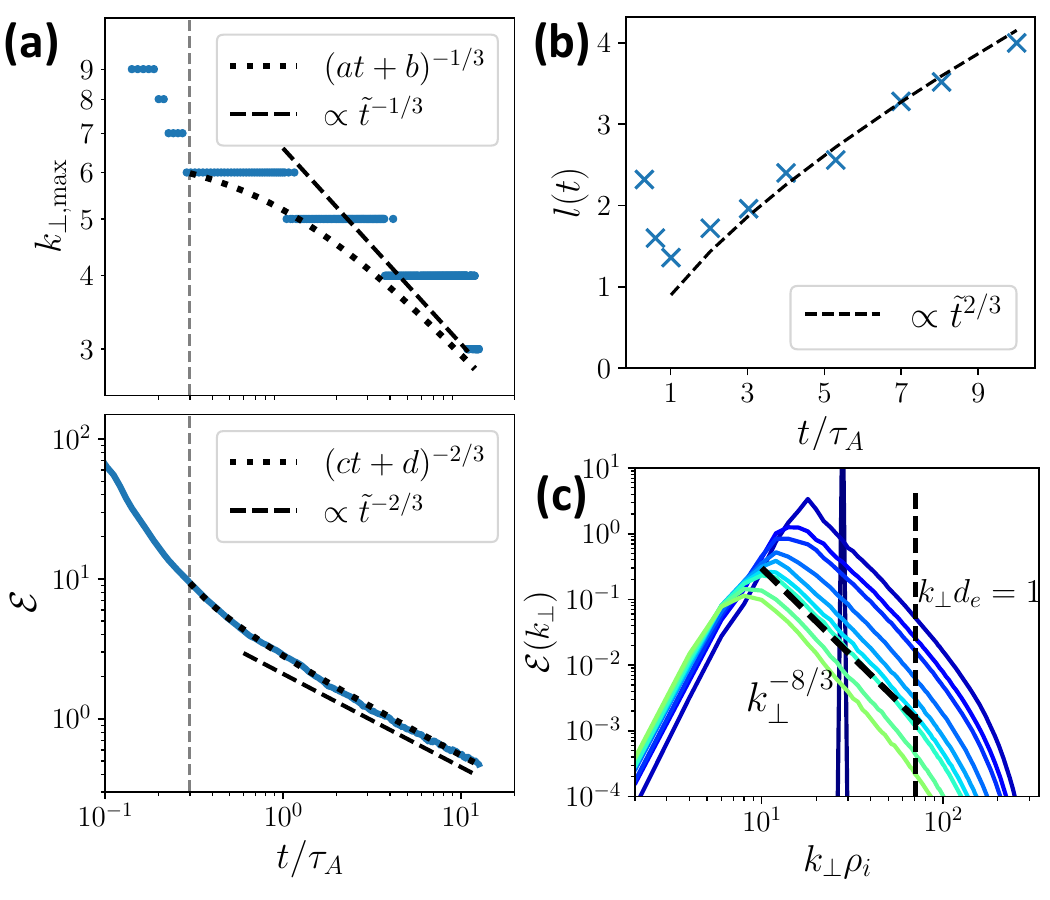}
\caption{\label{fig:2.3d} Three-dimensional simulation results. (a) Temporal evolution of the wavenumber corresponding to the peak of the magnetic spectrum (top panel) and total magnetic energy (bottom panel). $a,b,c,d$ are fitting parameters. We use the data to the right of the gray vertical dashed lines to avoid the transient stage. (b) Parallel coherence length $l$ at different times. (c) The perpendicular magnetic spectra at progressively later times (from blue to green). The vertical dashed line denotes the $d_e$ scale.}
\end{figure}

Fig.~\ref{fig:2.3d} presents the temporal evolution of some key quantities.
The top panel of Fig.~\ref{fig:2.3d} (a) shows the time evolution of the wavenumber at which the magnetic energy spectrum peaks, $k_{\perp,\rm{max}}$; whereas the bottom panel shows the box-averaged magnetic energy, $\mathcal{E}$. 
Both quantities conform to the theoretical predictions after $t\approx 0.3 \tau_A$, with the scaling laws of Eq.~\eqref{eq:scale_subrhoi} becoming applicable for $t \gtrsim 1.0\, \tau_A$~\footnote{The fit is slightly worse for $k_{\perp,\rm{max}}$ than for $\mathcal{E}$ due to the discrete nature of this diagnostic and the limited change in the value of $k_{\perp,\rm{max}}$, given the constrained size of the simulation domain.}.
Likewise, the parallel coherence length $l$, estimated using the five-point second-order structure function~\citep{cho2009simulations,cerri2019kinetic,zhou2023spectrum} (see the Supplemental Material for further details), aligns with Eq.~\eqref{eq:length} after approximately $t\approx 2.0\,\tau_A$, as evident from Fig.~\ref{fig:2.3d} (b)~\footnote{Before the system settles into a natural state of decaying turbulence described by our theory, it undergoes a transient stage characterized by the development of strong turbulence.
An essential indicator of strong turbulence is the establishment of critical balance, where flux tubes initially confined to box length break in the $z$-direction. This reduction in the parallel coherence length $l$ is captured by the first two data points in Fig.~\ref{fig:2.3d} (b).}.
We find (see Supplemental Material) that these temporal power laws also hold in the non-isothermal limit.

Fig.~\ref{fig:2.3d} (c) shows the evolution of the perpendicular magnetic-energy spectrum.
A reasonable agreement with a $k_{\perp}^{-8/3}$~\cite{boldyrev2012spectrum,zhou2023spectrum} inertial-range power-law behavior is observed, suggesting that, in addition to the inverse cascade arising from the flux-tube merger (which accounts for the shift of the peak of the magnetic spectrum to smaller values of $k_\perp$), there is also a direct cascade driven by the nonlinear interaction of kinetic Alfv\'en wave packets maintaining the $-8/3$ spectrum (see Section 3 of the Supplemental Material for energy transfer diagnostics).
This observation is consistent with measurements of turbulence in the Earth's magnetosheath in instances where electron-only reconnection is prevalent~\cite{stawarz2019properties}.

The similarity between the reconnection and the eddy turn-over times implies that, in our simulations, the inverse and forward energy cascades should indeed be comparable (see Section 3 of the Supplemental Material for evidence). However, we note that in systems where magnetic islands are not as volume-filling as in our numerical setup, the inverse cascade would not be expected to be energetically significant, and the forward cascade should dominate.

\paragraph{Transition to MHD scales.} As the flux tube (or island) merging proceeds, structures get progressively larger and eventually transition to MHD scales, i.e., $R \gg \rho_i$.
In this regime, electron-only reconnection no longer applies; instead, we expect the (normalized) reconnection rate to be a constant, independent of either the island size or the kinetic scales (Fig.~\ref{fig:1.rec_rate} indicates $\mathcal{R}_{\rm rec} \approx 0.2$).
This observation, together with flux conservation, is all that is required to conclude that, at this stage, the evolution scalings of resistive MHD~\cite{zhou2019magnetic,zhou2020multi,zhou2021statistical} should apply here as well. Namely, we expect
\begin{equation}
k_{\perp} \propto \tilde{t}^{-1 / 2}, \quad B\propto\tilde{t}^{-1 / 2}, \quad \mathcal{E}\propto \tilde{t}^{-1}.
\label{eq:scale_mhd}
\end{equation} 

To directly investigate these predictions, we conducted a two-dimensional simulation for which the initial condition is 
$\hat{A_z}(\mathbf{k_{\perp}}) = A_{z0}(k_{\perp}^4/k_c^4) \exp(i\phi(\mathbf{k_{\perp}}))$ for $k_{\perp} \leq k_c$ and  $\hat{A_z}(\mathbf{k_{\perp}}) = A_{z0}(k_{\perp}^4/k_c^4)\exp(1-k_{\perp}^4/k_c^4)\exp(i\phi(\mathbf{k_{\perp}}))$ for $k_{\perp} > k_c$, where $\phi(\mathbf{k_{\perp}})$ is a random phase, $A_{z0}=0.001$ and $k_c = 160$.
The initial magnetic spectrum is still a very narrow spike at very small scales (see the dark blue spike in Fig.~\ref{fig:2.3d} (c).
The domain is a doubly periodic box of dimensions $L_x \times L_y$, with $L_x=L_y= 2\pi$, and the spatial grid size is $8192^2$.
We set $\rho_i = 0.025$ ($\rho_s \approx 0.0177$) and $d_e= 0.002$, so that the initial condition satisfies $\rho_i^{-1}<k_{c}<d_e^{-1}$; thus, this simulation should first capture merging at kinetic scales (as in the 3D simulation discussed earlier), followed by a transition to MHD scales.

\begin{figure}[!htbp]
\includegraphics[width=0.5\textwidth]{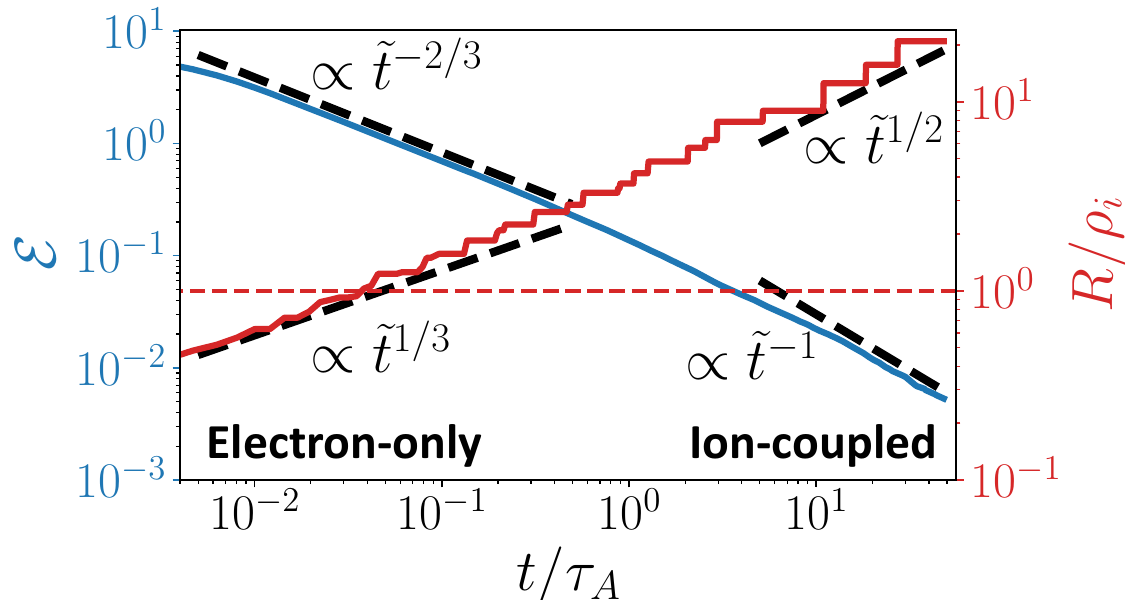}
\caption{\label{fig:3.2d} Temporal evolution of magnetic energy $\mathcal{E}$ (blue) and characteristic island radius $R$ (red) from the two-dimensional simulation. The horizontal red dashed line denotes $R=\rho_i$.}
\end{figure}

Time traces of magnetic energy $\mathcal{E}$ and characteristic radius $R$ of the magnetic structures, defined as $R \equiv 2\pi/4k_{\perp,{\rm max}}$, are presented in Fig.~\ref{fig:3.2d}.
Initially, these quantities are found to evolve according to the power laws  $\tilde{t}^{-2/3}$ and $\tilde{t}^{1/3}$, respectively, in accordance with Eq.~\eqref{eq:scale_subrhoi}.
As $R$ exceeds $\rho_i$ (around $t \approx 0.1 \tau_A$) the evolution of these quantities starts to deviate from those power laws.
This threshold is consistent with the departure from the electron-only reconnection regime (i.e., Eq.~\eqref{eq:rec_rate}) as $\rho_i/R$ decreases identified in Fig.~\ref{fig:1.rec_rate}.
Around $t\approx 10.0 \tau_A$, $R$ reaches $10\rho_i$, the scale at which the reconnection rate converges to the constant large-scale value and ions are fully coupled (see Fig.~\ref{fig:1.rec_rate}).
Consistently, transitions to $\mathcal{E}\sim\tilde{t}^{-1}$  and $R\sim\tilde{t}^{1/2}$ power-law scalings are observed, in agreement with the predictions of Eq.~\eqref{eq:scale_mhd}.

\paragraph{Conclusions.}
We present an analytical model for electron-only strong-guide-field reconnection in low-$\beta$ plasmas. 
Unlike large-scale collisionless reconnection, where the (normalized) reconnection rate is conjectured to be a constant of order $0.1$~\cite{cassak2017review,liu2017does}, we find that this regime is characterized by reconnection rates that scale as the ratio between the ion gyroradius and the size of the reconnecting structure and which can significantly exceed 0.1. 
This finding provides a theoretical explanation for previous observational and numerical studies~\citep[e.g.,][]{sharma2019transition,greess2022kinetic,guan2023reconnection}.
The validity of our model is confirmed through direct numerical simulations, which also reveal that deviation from the fully ion-coupled regime begins when $
R \approx 10\rho_i$, consistent with previous numerical studies~\cite{sharma2019transition} in a different plasma regime.
Although our focus in this study is low beta plasmas, we also propose an extension of our derivation to  $\beta_e \sim 1$ (see Supplemental Material) which appears to be consistent with existing numerical data.

Using this reconnection model, we derive time-dependent scaling laws for inverse magnetic-energy transfer in the sub-$\rho_i$ range, which we validate via a three-dimensional direct numerical simulation. 
Finally, through a two-dimensional simulation starting with random Gaussian magnetic fields, we demonstrate the ability of magnetic-island mergers to transport magnetic energy from kinetic to MHD scales, accompanied by predicted changes in the time-dependent scaling laws.
In addition to its relevance to the understanding of the self-organization of systems dominated by flux-rope-like structures~\citep[e.g.,][]{eastwood2016ion,sun2019mms,huang2021electron},
this conclusion lends credence to recently proposed ideas for the viability of electron-scale seeds as the origin of large-scale magnetic fields~\cite{zhou2020multi,zhou2022spontaneous,zhou2023magnetogenesis}.

\paragraph{Acknowledgements.} This work was supported by DOE Awards DE-SC0022012, DE-FG02-91-ER54109, and by the Schmidt Futures at the Institute for Advanced Study.
This work used resources of the Satori cluster at the MGHPCC facility funded by DOE award No. DE-FG02-91-ER54109, and the National Energy Research Scientific Computing Center, a DOE Office of Science User Facility supported by the Office of Science of the U.S. Department of Energy under Contract No. DE-AC02-05CH11231 using NERSC award FES-ERCAP0020063.

\appendix
\section{Further details of reconnection rate derivation}

\begin{figure}[!htbp]
\includegraphics[width=0.4\textwidth]{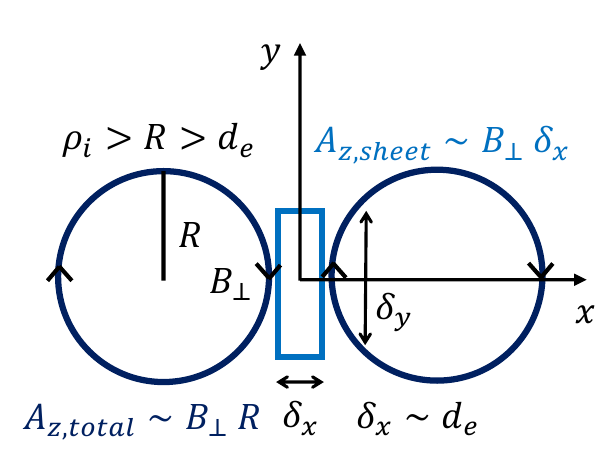}
\caption{\label{app:fig:1.rec} A schematic of electron-only reconnection configuration. Two magnetic islands (represented by dark blue circles) or radius $R$ such that $d_e<R<\rho_i$ and magnetic field $B_\perp$ reconnect with each other via a current sheet (central blue rectangle) of dimensions $\delta_x\sim d_e$ and $\delta_y\sim R$.}
\end{figure}

A schematic of the reconnection configuration under consideration is shown in Fig.~\ref{app:fig:1.rec}. The key quantities associated with the magnetic islands are the radius $R$, the reconnecting magnetic field $B_{\perp}$, and the total magnetic flux $A_{z, \rm{total}} = B_{\perp} R$. For the current sheet, the relevant quantities are its width $\delta_x \sim d_e$, its length $\delta_y \sim R$, the reconnecting flux $A_{z, \rm{sheet}} = B_{\perp} \delta_x$, and the out-of-plane current $\nabla_{\perp}^2 A_{z, \rm{sheet}} = -(4\pi/c)J_z \sim B_{\perp}/\delta_x$ (derived from Ampere's law).

We start by rewriting Eqs.~(\ref{eq:ne}-\ref{eq:gk_poisson}) in their 2D form:
\begin{equation}
    \frac{\partial \delta n_e / n_{0 e}}{\partial t}+\frac{c}{B_0}\left\{\varphi, \delta n_e / n_{0 e}\right\}  =\frac{1}{B_0}\left\{A_{\|}, \frac{e}{c m_e} d_e^2 \nabla_{\perp}^2 A_{\|}\right\}, \label{app:eq:ne_1}
\end{equation}
\begin{equation}
\begin{aligned}
    &\frac{\partial\left(A_{\|}- d_e^2 \nabla_{\perp}^2 A_{\|}\right)}{\partial t}+\frac{c}{B_0}\left\{\varphi, A_{\|}-d_e^2 \nabla_{\perp}^2 A_{\|}\right\}  \\
    =&-\frac{c T_{0 e}}{e} \frac{1}{B_0}\left\{A_{\|}, \frac{\delta n_e}{n_{0 e}}\right\}
\end{aligned}
\label{app:eq:Az_1}
\end{equation}
\begin{equation}
        \frac{\delta n_e}{n_{0e}}  =\frac{T_{0e}}{T_{0i}}(\hat{\Gamma}_0-1)\frac{e\varphi}{T_{0e}}
    \label{app:eq:poisson_1}.
\end{equation}
In the large $k_{\perp}\rho_i$ limit, Eq.~\eqref{app:eq:poisson_1} reduces to~\citep{zocco2011reduced}
\begin{equation}
        \frac{\delta n_e}{n_{0e}} =-\frac{e\varphi}{T_{0i}}.
        \label{app:eq:poisson_reduced}
\end{equation}

Next, we balance the terms in the above equations within the current sheet, where the spatial derivatives are approximated as $\partial_x \sim 1/\delta_x \sim 1/d_e$ and $\partial_y \sim 1/\delta_y \sim 1/R$. We introduce $\tau$, the time taken to reconnect the flux within the current sheet, represented by $A_{z, \rm{sheet}}$.

In Eq.~\eqref{app:eq:ne_1}, the Poisson bracket between $\varphi$ and $\delta n_e$ vanishes due to Eq.~\eqref{app:eq:poisson_reduced}.
The remaining terms are estimated as:
\begin{equation}
\begin{aligned}
    \frac{\delta n_e/n_{0e}}{\tau} 
    &\sim \frac{e d_e^2}{B_0 cm_e}\partial_x A_{z, \rm{sheet}}\partial_y \nabla_{\perp}^2 A_{z, \rm{sheet}} \\
    &\sim \frac{e d_e^2}{B_0 cm_e} B_{\perp} \frac{B_{\perp}/\delta_x}{R}.
\end{aligned}    
\label{app:eq:ne_2}
\end{equation}

In Eq.~\eqref{app:eq:Az_1}, since $d_e^2 \nabla_{\perp}^2A_{z, \rm{sheet}} \sim d_e^2B_{\perp}/\delta_x \sim (d_e/\delta_x)^2 A_{z, \rm{sheet}} \sim A_{z, \rm{sheet}}$, the electron inertial term $d_e^2 \nabla_{\perp}^2 A_{z, \rm{sheet}}$ does not affect term balancing. The other terms are estimated as
\begin{equation}
\begin{aligned}
    \frac{A_{z, \rm{sheet}}}{\tau} &\sim \frac{c(T_{0e}+T_{0i})}{eB_0} \partial_x A_{z, \rm{sheet}} \partial_y (\delta n_e/n_{0e}) \\
    &\sim \frac{c(T_{0e}+T_{0i})}{eB_0} B_{\perp} \frac{\delta n_e/n_{0e}}{R}, \label{app:eq:Az_2}
\end{aligned}
\end{equation}
noting that the two Poisson brackets combine using Eq.~\eqref{app:eq:poisson_reduced}.

Combining Eq.~\eqref{app:eq:ne_2} and Eq.~\eqref{app:eq:Az_2} to eliminate $\delta n_e/n_{0e}$, we derive
\begin{equation}
    \tau \sim \frac{1}{\sqrt{1+T_{0i}/T_{0e}}}\frac{R}{\rho_s} \frac{d_e}{v_{A,\perp}}, \label{app:eq:tau}
\end{equation}
where $v_{A,\perp} = B_{\perp}/\sqrt{4\pi n_0 m_i}$.

Finally, the reconnection time $\tau_{\rm rec}$, representing the time taken to reconnect the entire magnetic flux, relates to $\tau$ via $\tau_{\rm rec}/\tau \sim A_{z, \rm{total}}/A_{z, \rm{sheet}}$. 
Thus, we obtain: 
\begin{equation} \tau_{\rm rec} \sim \frac{1}{\sqrt{1+T_{0i}/T_{0e}}}\frac{R}{\rho_s} \frac{R}{v_{A,\perp}}. 
\end{equation}

\nocite{*}
\bibliography{main}

\begin{thebibliography}{84}%
\makeatletter
\providecommand \@ifxundefined [1]{%
 \@ifx{#1\undefined}
}%
\providecommand \@ifnum [1]{%
 \ifnum #1\expandafter \@firstoftwo
 \else \expandafter \@secondoftwo
 \fi
}%
\providecommand \@ifx [1]{%
 \ifx #1\expandafter \@firstoftwo
 \else \expandafter \@secondoftwo
 \fi
}%
\providecommand \natexlab [1]{#1}%
\providecommand \enquote  [1]{``#1''}%
\providecommand \bibnamefont  [1]{#1}%
\providecommand \bibfnamefont [1]{#1}%
\providecommand \citenamefont [1]{#1}%
\providecommand \href@noop [0]{\@secondoftwo}%
\providecommand \href [0]{\begingroup \@sanitize@url \@href}%
\providecommand \@href[1]{\@@startlink{#1}\@@href}%
\providecommand \@@href[1]{\endgroup#1\@@endlink}%
\providecommand \@sanitize@url [0]{\catcode `\\12\catcode `\$12\catcode `\&12\catcode `\#12\catcode `\^12\catcode `\_12\catcode `\%12\relax}%
\providecommand \@@startlink[1]{}%
\providecommand \@@endlink[0]{}%
\providecommand \url  [0]{\begingroup\@sanitize@url \@url }%
\providecommand \@url [1]{\endgroup\@href {#1}{\urlprefix }}%
\providecommand \urlprefix  [0]{URL }%
\providecommand \Eprint [0]{\href }%
\providecommand \doibase [0]{https://doi.org/}%
\providecommand \selectlanguage [0]{\@gobble}%
\providecommand \bibinfo  [0]{\@secondoftwo}%
\providecommand \bibfield  [0]{\@secondoftwo}%
\providecommand \translation [1]{[#1]}%
\providecommand \BibitemOpen [0]{}%
\providecommand \bibitemStop [0]{}%
\providecommand \bibitemNoStop [0]{.\EOS\space}%
\providecommand \EOS [0]{\spacefactor3000\relax}%
\providecommand \BibitemShut  [1]{\csname bibitem#1\endcsname}%
\let\auto@bib@innerbib\@empty
\bibitem [{\citenamefont {Banerjee}\ and\ \citenamefont {Jedamzik}(2004)}]{banerjee2004evolution}%
  \BibitemOpen
  \bibfield  {author} {\bibinfo {author} {\bibfnamefont {R.}~\bibnamefont {Banerjee}}\ and\ \bibinfo {author} {\bibfnamefont {K.}~\bibnamefont {Jedamzik}},\ }\bibfield  {title} {\bibinfo {title} {Evolution of cosmic magnetic fields: From the very early universe, to recombination, to the present},\ }\href {https://doi.org/10.1103/PhysRevD.70.123003} {\bibfield  {journal} {\bibinfo  {journal} {Phys. Rev. D}\ }\textbf {\bibinfo {volume} {70}},\ \bibinfo {pages} {123003} (\bibinfo {year} {2004})}\BibitemShut {NoStop}%
\bibitem [{\citenamefont {Durrer}\ and\ \citenamefont {Neronov}(2013)}]{durrer2013cosmological}%
  \BibitemOpen
  \bibfield  {author} {\bibinfo {author} {\bibfnamefont {R.}~\bibnamefont {Durrer}}\ and\ \bibinfo {author} {\bibfnamefont {A.}~\bibnamefont {Neronov}},\ }\bibfield  {title} {\bibinfo {title} {Cosmological magnetic fields: their generation, evolution and observation},\ }\href@noop {} {\bibfield  {journal} {\bibinfo  {journal} {The Astronomy and Astrophysics Review}\ }\textbf {\bibinfo {volume} {21}},\ \bibinfo {pages} {1} (\bibinfo {year} {2013})}\BibitemShut {NoStop}%
\bibitem [{\citenamefont {Hosking}\ and\ \citenamefont {Schekochihin}(2023)}]{hosking2023cosmic}%
  \BibitemOpen
  \bibfield  {author} {\bibinfo {author} {\bibfnamefont {D.~N.}\ \bibnamefont {Hosking}}\ and\ \bibinfo {author} {\bibfnamefont {A.~A.}\ \bibnamefont {Schekochihin}},\ }\bibfield  {title} {\bibinfo {title} {Cosmic-void observations reconciled with primordial magnetogenesis},\ }\href@noop {} {\bibfield  {journal} {\bibinfo  {journal} {Nature Communications}\ }\textbf {\bibinfo {volume} {14}},\ \bibinfo {pages} {7523} (\bibinfo {year} {2023})}\BibitemShut {NoStop}%
\bibitem [{\citenamefont {Chen}\ \emph {et~al.}(2011)\citenamefont {Chen}, \citenamefont {Mallet}, \citenamefont {Yousef}, \citenamefont {Schekochihin},\ and\ \citenamefont {Horbury}}]{chen2011anisotropy}%
  \BibitemOpen
  \bibfield  {author} {\bibinfo {author} {\bibfnamefont {C.}~\bibnamefont {Chen}}, \bibinfo {author} {\bibfnamefont {A.}~\bibnamefont {Mallet}}, \bibinfo {author} {\bibfnamefont {T.}~\bibnamefont {Yousef}}, \bibinfo {author} {\bibfnamefont {A.}~\bibnamefont {Schekochihin}},\ and\ \bibinfo {author} {\bibfnamefont {T.}~\bibnamefont {Horbury}},\ }\bibfield  {title} {\bibinfo {title} {Anisotropy of {A}lfv{\'e}nic turbulence in the solar wind and numerical simulations},\ }\href@noop {} {\bibfield  {journal} {\bibinfo  {journal} {Monthly Notices of the Royal Astronomical Society}\ }\textbf {\bibinfo {volume} {415}},\ \bibinfo {pages} {3219} (\bibinfo {year} {2011})}\BibitemShut {NoStop}%
\bibitem [{\citenamefont {Khabarova}\ \emph {et~al.}(2015)\citenamefont {Khabarova}, \citenamefont {Zank}, \citenamefont {Li}, \citenamefont {Le~Roux}, \citenamefont {Webb}, \citenamefont {Dosch},\ and\ \citenamefont {Malandraki}}]{khabarova2015small}%
  \BibitemOpen
  \bibfield  {author} {\bibinfo {author} {\bibfnamefont {O.}~\bibnamefont {Khabarova}}, \bibinfo {author} {\bibfnamefont {G.}~\bibnamefont {Zank}}, \bibinfo {author} {\bibfnamefont {G.}~\bibnamefont {Li}}, \bibinfo {author} {\bibfnamefont {J.}~\bibnamefont {Le~Roux}}, \bibinfo {author} {\bibfnamefont {G.}~\bibnamefont {Webb}}, \bibinfo {author} {\bibfnamefont {A.}~\bibnamefont {Dosch}},\ and\ \bibinfo {author} {\bibfnamefont {O.}~\bibnamefont {Malandraki}},\ }\bibfield  {title} {\bibinfo {title} {Small-scale magnetic islands in the solar wind and their role in particle acceleration. i. dynamics of magnetic islands near the heliospheric current sheet},\ }\href@noop {} {\bibfield  {journal} {\bibinfo  {journal} {The Astrophysical Journal}\ }\textbf {\bibinfo {volume} {808}},\ \bibinfo {pages} {181} (\bibinfo {year} {2015})}\BibitemShut {NoStop}%
\bibitem [{\citenamefont {Drake}\ \emph {et~al.}(2012)\citenamefont {Drake}, \citenamefont {Swisdak},\ and\ \citenamefont {Fermo}}]{drake2012power}%
  \BibitemOpen
  \bibfield  {author} {\bibinfo {author} {\bibfnamefont {J.}~\bibnamefont {Drake}}, \bibinfo {author} {\bibfnamefont {M.}~\bibnamefont {Swisdak}},\ and\ \bibinfo {author} {\bibfnamefont {R.}~\bibnamefont {Fermo}},\ }\bibfield  {title} {\bibinfo {title} {The power-law spectra of energetic particles during multi-island magnetic reconnection},\ }\href@noop {} {\bibfield  {journal} {\bibinfo  {journal} {The Astrophysical Journal Letters}\ }\textbf {\bibinfo {volume} {763}},\ \bibinfo {pages} {L5} (\bibinfo {year} {2012})}\BibitemShut {NoStop}%
\bibitem [{\citenamefont {Pouquet}(1978)}]{pouquet1978two}%
  \BibitemOpen
  \bibfield  {author} {\bibinfo {author} {\bibfnamefont {A.}~\bibnamefont {Pouquet}},\ }\bibfield  {title} {\bibinfo {title} {On two-dimensional magnetohydrodynamic turbulence},\ }\href@noop {} {\bibfield  {journal} {\bibinfo  {journal} {Journal of Fluid Mechanics}\ }\textbf {\bibinfo {volume} {88}},\ \bibinfo {pages} {1} (\bibinfo {year} {1978})}\BibitemShut {NoStop}%
\bibitem [{\citenamefont {Matthaeus}\ and\ \citenamefont {Lamkin}(1986)}]{matthaeus1986turbulent}%
  \BibitemOpen
  \bibfield  {author} {\bibinfo {author} {\bibfnamefont {W.}~\bibnamefont {Matthaeus}}\ and\ \bibinfo {author} {\bibfnamefont {S.~L.}\ \bibnamefont {Lamkin}},\ }\bibfield  {title} {\bibinfo {title} {Turbulent magnetic reconnection},\ }\href@noop {} {\bibfield  {journal} {\bibinfo  {journal} {The Physics of fluids}\ }\textbf {\bibinfo {volume} {29}},\ \bibinfo {pages} {2513} (\bibinfo {year} {1986})}\BibitemShut {NoStop}%
\bibitem [{\citenamefont {Christensson}\ \emph {et~al.}(2001)\citenamefont {Christensson}, \citenamefont {Hindmarsh},\ and\ \citenamefont {Brandenburg}}]{christensson2001inverse}%
  \BibitemOpen
  \bibfield  {author} {\bibinfo {author} {\bibfnamefont {M.}~\bibnamefont {Christensson}}, \bibinfo {author} {\bibfnamefont {M.}~\bibnamefont {Hindmarsh}},\ and\ \bibinfo {author} {\bibfnamefont {A.}~\bibnamefont {Brandenburg}},\ }\bibfield  {title} {\bibinfo {title} {Inverse cascade in decaying three-dimensional magnetohydrodynamic turbulence},\ }\href@noop {} {\bibfield  {journal} {\bibinfo  {journal} {Physical Review E}\ }\textbf {\bibinfo {volume} {64}},\ \bibinfo {pages} {056405} (\bibinfo {year} {2001})}\BibitemShut {NoStop}%
\bibitem [{\citenamefont {Brandenburg}\ \emph {et~al.}(2015)\citenamefont {Brandenburg}, \citenamefont {Kahniashvili},\ and\ \citenamefont {Tevzadze}}]{brandenburg2015nonhelical}%
  \BibitemOpen
  \bibfield  {author} {\bibinfo {author} {\bibfnamefont {A.}~\bibnamefont {Brandenburg}}, \bibinfo {author} {\bibfnamefont {T.}~\bibnamefont {Kahniashvili}},\ and\ \bibinfo {author} {\bibfnamefont {A.~G.}\ \bibnamefont {Tevzadze}},\ }\bibfield  {title} {\bibinfo {title} {Nonhelical inverse transfer of a decaying turbulent magnetic field},\ }\href@noop {} {\bibfield  {journal} {\bibinfo  {journal} {Physical review letters}\ }\textbf {\bibinfo {volume} {114}},\ \bibinfo {pages} {075001} (\bibinfo {year} {2015})}\BibitemShut {NoStop}%
\bibitem [{\citenamefont {Reppin}\ and\ \citenamefont {Banerjee}(2017)}]{reppin2017nonhelical}%
  \BibitemOpen
  \bibfield  {author} {\bibinfo {author} {\bibfnamefont {J.}~\bibnamefont {Reppin}}\ and\ \bibinfo {author} {\bibfnamefont {R.}~\bibnamefont {Banerjee}},\ }\bibfield  {title} {\bibinfo {title} {Nonhelical turbulence and the inverse transfer of energy: A parameter study},\ }\href@noop {} {\bibfield  {journal} {\bibinfo  {journal} {Physical Review E}\ }\textbf {\bibinfo {volume} {96}},\ \bibinfo {pages} {053105} (\bibinfo {year} {2017})}\BibitemShut {NoStop}%
\bibitem [{\citenamefont {Zhou}\ \emph {et~al.}(2019)\citenamefont {Zhou}, \citenamefont {Bhat}, \citenamefont {Loureiro},\ and\ \citenamefont {Uzdensky}}]{zhou2019magnetic}%
  \BibitemOpen
  \bibfield  {author} {\bibinfo {author} {\bibfnamefont {M.}~\bibnamefont {Zhou}}, \bibinfo {author} {\bibfnamefont {P.}~\bibnamefont {Bhat}}, \bibinfo {author} {\bibfnamefont {N.~F.}\ \bibnamefont {Loureiro}},\ and\ \bibinfo {author} {\bibfnamefont {D.~A.}\ \bibnamefont {Uzdensky}},\ }\bibfield  {title} {\bibinfo {title} {Magnetic island merger as a mechanism for inverse magnetic energy transfer},\ }\href@noop {} {\bibfield  {journal} {\bibinfo  {journal} {Physical Review Research}\ }\textbf {\bibinfo {volume} {1}},\ \bibinfo {pages} {012004} (\bibinfo {year} {2019})}\BibitemShut {NoStop}%
\bibitem [{\citenamefont {Zhou}\ \emph {et~al.}(2020)\citenamefont {Zhou}, \citenamefont {Loureiro},\ and\ \citenamefont {Uzdensky}}]{zhou2020multi}%
  \BibitemOpen
  \bibfield  {author} {\bibinfo {author} {\bibfnamefont {M.}~\bibnamefont {Zhou}}, \bibinfo {author} {\bibfnamefont {N.~F.}\ \bibnamefont {Loureiro}},\ and\ \bibinfo {author} {\bibfnamefont {D.~A.}\ \bibnamefont {Uzdensky}},\ }\bibfield  {title} {\bibinfo {title} {Multi-scale dynamics of magnetic flux tubes and inverse magnetic energy transfer},\ }\href@noop {} {\bibfield  {journal} {\bibinfo  {journal} {Journal of Plasma Physics}\ }\textbf {\bibinfo {volume} {86}},\ \bibinfo {pages} {535860401} (\bibinfo {year} {2020})}\BibitemShut {NoStop}%
\bibitem [{\citenamefont {Bhat}\ \emph {et~al.}(2021)\citenamefont {Bhat}, \citenamefont {Zhou},\ and\ \citenamefont {Loureiro}}]{bhat2021inverse}%
  \BibitemOpen
  \bibfield  {author} {\bibinfo {author} {\bibfnamefont {P.}~\bibnamefont {Bhat}}, \bibinfo {author} {\bibfnamefont {M.}~\bibnamefont {Zhou}},\ and\ \bibinfo {author} {\bibfnamefont {N.~F.}\ \bibnamefont {Loureiro}},\ }\bibfield  {title} {\bibinfo {title} {Inverse energy transfer in decaying, three-dimensional, non-helical magnetic turbulence due to magnetic reconnection},\ }\href@noop {} {\bibfield  {journal} {\bibinfo  {journal} {Monthly Notices of the Royal Astronomical Society}\ }\textbf {\bibinfo {volume} {501}},\ \bibinfo {pages} {3074} (\bibinfo {year} {2021})}\BibitemShut {NoStop}%
\bibitem [{\citenamefont {Zhou}\ \emph {et~al.}(2021)\citenamefont {Zhou}, \citenamefont {Wu}, \citenamefont {Loureiro},\ and\ \citenamefont {Uzdensky}}]{zhou2021statistical}%
  \BibitemOpen
  \bibfield  {author} {\bibinfo {author} {\bibfnamefont {M.}~\bibnamefont {Zhou}}, \bibinfo {author} {\bibfnamefont {D.~H.}\ \bibnamefont {Wu}}, \bibinfo {author} {\bibfnamefont {N.~F.}\ \bibnamefont {Loureiro}},\ and\ \bibinfo {author} {\bibfnamefont {D.~A.}\ \bibnamefont {Uzdensky}},\ }\bibfield  {title} {\bibinfo {title} {Statistical description of coalescing magnetic islands via magnetic reconnection},\ }\href@noop {} {\bibfield  {journal} {\bibinfo  {journal} {Journal of Plasma Physics}\ }\textbf {\bibinfo {volume} {87}},\ \bibinfo {pages} {905870620} (\bibinfo {year} {2021})}\BibitemShut {NoStop}%
\bibitem [{\citenamefont {Hosking}\ and\ \citenamefont {Schekochihin}(2021)}]{hosking2021reconnection}%
  \BibitemOpen
  \bibfield  {author} {\bibinfo {author} {\bibfnamefont {D.~N.}\ \bibnamefont {Hosking}}\ and\ \bibinfo {author} {\bibfnamefont {A.~A.}\ \bibnamefont {Schekochihin}},\ }\bibfield  {title} {\bibinfo {title} {Reconnection-controlled decay of magnetohydrodynamic turbulence and the role of invariants},\ }\href@noop {} {\bibfield  {journal} {\bibinfo  {journal} {Physical Review X}\ }\textbf {\bibinfo {volume} {11}},\ \bibinfo {pages} {041005} (\bibinfo {year} {2021})}\BibitemShut {NoStop}%
\bibitem [{\citenamefont {Milanese}(2023)}]{milanese2023}%
  \BibitemOpen
  \bibfield  {author} {\bibinfo {author} {\bibfnamefont {L.~M.}\ \bibnamefont {Milanese}},\ }\emph {\bibinfo {title} {Turbulent Dynamics Under Two Ideal Invariants: Dynamic Phase Alignment in Plasmas and Nonionized Fluids}},\ \href@noop {} {\bibinfo {type} {Phd thesis}},\ \bibinfo  {school} {Massachusetts Institute of Technology}, \bibinfo {address} {Cambridge, MA} (\bibinfo {year} {2023}),\ \bibinfo {note} {available at \url{https://hdl.handle.net/1721.1/151221}}\BibitemShut {NoStop}%
\bibitem [{\citenamefont {Stawarz}\ \emph {et~al.}(2018)\citenamefont {Stawarz}, \citenamefont {Eastwood}, \citenamefont {Genestreti}, \citenamefont {Nakamura}, \citenamefont {Ergun}, \citenamefont {Burgess}, \citenamefont {Burch}, \citenamefont {Fuselier}, \citenamefont {Gershman}, \citenamefont {Giles} \emph {et~al.}}]{stawarz2018intense}%
  \BibitemOpen
  \bibfield  {author} {\bibinfo {author} {\bibfnamefont {J.}~\bibnamefont {Stawarz}}, \bibinfo {author} {\bibfnamefont {J.}~\bibnamefont {Eastwood}}, \bibinfo {author} {\bibfnamefont {K.}~\bibnamefont {Genestreti}}, \bibinfo {author} {\bibfnamefont {R.}~\bibnamefont {Nakamura}}, \bibinfo {author} {\bibfnamefont {R.}~\bibnamefont {Ergun}}, \bibinfo {author} {\bibfnamefont {D.}~\bibnamefont {Burgess}}, \bibinfo {author} {\bibfnamefont {J.}~\bibnamefont {Burch}}, \bibinfo {author} {\bibfnamefont {S.}~\bibnamefont {Fuselier}}, \bibinfo {author} {\bibfnamefont {D.}~\bibnamefont {Gershman}}, \bibinfo {author} {\bibfnamefont {B.~L.}\ \bibnamefont {Giles}}, \emph {et~al.},\ }\bibfield  {title} {\bibinfo {title} {Intense electric fields and electron-scale substructure within magnetotail flux ropes as revealed by the magnetospheric multiscale mission},\ }\href@noop {} {\bibfield  {journal} {\bibinfo  {journal} {Geophysical Research Letters}\ }\textbf {\bibinfo {volume} {45}},\ \bibinfo {pages} {8783} (\bibinfo {year}
  {2018})}\BibitemShut {NoStop}%
\bibitem [{\citenamefont {Zhong}\ \emph {et~al.}(2020)\citenamefont {Zhong}, \citenamefont {Zhou}, \citenamefont {Tang}, \citenamefont {Deng}, \citenamefont {Turner}, \citenamefont {Cohen}, \citenamefont {Pang}, \citenamefont {Man}, \citenamefont {Russell}, \citenamefont {Giles} \emph {et~al.}}]{zhong2020direct}%
  \BibitemOpen
  \bibfield  {author} {\bibinfo {author} {\bibfnamefont {Z.}~\bibnamefont {Zhong}}, \bibinfo {author} {\bibfnamefont {M.}~\bibnamefont {Zhou}}, \bibinfo {author} {\bibfnamefont {R.}~\bibnamefont {Tang}}, \bibinfo {author} {\bibfnamefont {X.}~\bibnamefont {Deng}}, \bibinfo {author} {\bibfnamefont {D.}~\bibnamefont {Turner}}, \bibinfo {author} {\bibfnamefont {I.}~\bibnamefont {Cohen}}, \bibinfo {author} {\bibfnamefont {Y.}~\bibnamefont {Pang}}, \bibinfo {author} {\bibfnamefont {H.}~\bibnamefont {Man}}, \bibinfo {author} {\bibfnamefont {C.}~\bibnamefont {Russell}}, \bibinfo {author} {\bibfnamefont {B.}~\bibnamefont {Giles}}, \emph {et~al.},\ }\bibfield  {title} {\bibinfo {title} {Direct evidence for electron acceleration within ion-scale flux rope},\ }\href@noop {} {\bibfield  {journal} {\bibinfo  {journal} {Geophysical Research Letters}\ }\textbf {\bibinfo {volume} {47}},\ \bibinfo {pages} {e2019GL085141} (\bibinfo {year} {2020})}\BibitemShut {NoStop}%
\bibitem [{\citenamefont {Huang}\ \emph {et~al.}(2021)\citenamefont {Huang}, \citenamefont {Xiong}, \citenamefont {Song}, \citenamefont {Nan}, \citenamefont {Yuan}, \citenamefont {Jiang}, \citenamefont {Deng},\ and\ \citenamefont {Yu}}]{huang2021electron}%
  \BibitemOpen
  \bibfield  {author} {\bibinfo {author} {\bibfnamefont {S.}~\bibnamefont {Huang}}, \bibinfo {author} {\bibfnamefont {Q.}~\bibnamefont {Xiong}}, \bibinfo {author} {\bibfnamefont {L.}~\bibnamefont {Song}}, \bibinfo {author} {\bibfnamefont {J.}~\bibnamefont {Nan}}, \bibinfo {author} {\bibfnamefont {Z.}~\bibnamefont {Yuan}}, \bibinfo {author} {\bibfnamefont {K.}~\bibnamefont {Jiang}}, \bibinfo {author} {\bibfnamefont {X.}~\bibnamefont {Deng}},\ and\ \bibinfo {author} {\bibfnamefont {L.}~\bibnamefont {Yu}},\ }\bibfield  {title} {\bibinfo {title} {Electron-only reconnection in an ion-scale current sheet at the magnetopause},\ }\href@noop {} {\bibfield  {journal} {\bibinfo  {journal} {The Astrophysical Journal}\ }\textbf {\bibinfo {volume} {922}},\ \bibinfo {pages} {54} (\bibinfo {year} {2021})}\BibitemShut {NoStop}%
\bibitem [{\citenamefont {Hasegawa}\ \emph {et~al.}(2023)\citenamefont {Hasegawa}, \citenamefont {Denton}, \citenamefont {Dokgo}, \citenamefont {Hwang}, \citenamefont {Nakamura},\ and\ \citenamefont {Burch}}]{hasegawa2023ion}%
  \BibitemOpen
  \bibfield  {author} {\bibinfo {author} {\bibfnamefont {H.}~\bibnamefont {Hasegawa}}, \bibinfo {author} {\bibfnamefont {R.}~\bibnamefont {Denton}}, \bibinfo {author} {\bibfnamefont {K.}~\bibnamefont {Dokgo}}, \bibinfo {author} {\bibfnamefont {K.-J.}\ \bibnamefont {Hwang}}, \bibinfo {author} {\bibfnamefont {T.}~\bibnamefont {Nakamura}},\ and\ \bibinfo {author} {\bibfnamefont {J.}~\bibnamefont {Burch}},\ }\bibfield  {title} {\bibinfo {title} {Ion-scale magnetic flux rope generated from electron-scale magnetopause current sheet: Magnetospheric multiscale observations},\ }\href@noop {} {\bibfield  {journal} {\bibinfo  {journal} {Journal of Geophysical Research: Space Physics}\ }\textbf {\bibinfo {volume} {128}},\ \bibinfo {pages} {e2022JA031092} (\bibinfo {year} {2023})}\BibitemShut {NoStop}%
\bibitem [{\citenamefont {Phan}\ \emph {et~al.}(2018)\citenamefont {Phan}, \citenamefont {Eastwood}, \citenamefont {Shay}, \citenamefont {Drake}, \citenamefont {Sonnerup}, \citenamefont {Fujimoto}, \citenamefont {Cassak}, \citenamefont {{\O}ieroset}, \citenamefont {Burch}, \citenamefont {Torbert} \emph {et~al.}}]{phan2018electron}%
  \BibitemOpen
  \bibfield  {author} {\bibinfo {author} {\bibfnamefont {T.}~\bibnamefont {Phan}}, \bibinfo {author} {\bibfnamefont {J.~P.}\ \bibnamefont {Eastwood}}, \bibinfo {author} {\bibfnamefont {M.}~\bibnamefont {Shay}}, \bibinfo {author} {\bibfnamefont {J.}~\bibnamefont {Drake}}, \bibinfo {author} {\bibfnamefont {B.~{\"O}.}\ \bibnamefont {Sonnerup}}, \bibinfo {author} {\bibfnamefont {M.}~\bibnamefont {Fujimoto}}, \bibinfo {author} {\bibfnamefont {P.}~\bibnamefont {Cassak}}, \bibinfo {author} {\bibfnamefont {M.}~\bibnamefont {{\O}ieroset}}, \bibinfo {author} {\bibfnamefont {J.}~\bibnamefont {Burch}}, \bibinfo {author} {\bibfnamefont {R.}~\bibnamefont {Torbert}}, \emph {et~al.},\ }\bibfield  {title} {\bibinfo {title} {Electron magnetic reconnection without ion coupling in earth’s turbulent magnetosheath},\ }\href@noop {} {\bibfield  {journal} {\bibinfo  {journal} {Nature}\ }\textbf {\bibinfo {volume} {557}},\ \bibinfo {pages} {202} (\bibinfo {year} {2018})}\BibitemShut {NoStop}%
\bibitem [{\citenamefont {Stawarz}\ \emph {et~al.}(2019)\citenamefont {Stawarz}, \citenamefont {Eastwood}, \citenamefont {Phan}, \citenamefont {Gingell}, \citenamefont {Shay}, \citenamefont {Burch}, \citenamefont {Ergun}, \citenamefont {Giles}, \citenamefont {Gershman}, \citenamefont {Le~Contel} \emph {et~al.}}]{stawarz2019properties}%
  \BibitemOpen
  \bibfield  {author} {\bibinfo {author} {\bibfnamefont {J.~E.}\ \bibnamefont {Stawarz}}, \bibinfo {author} {\bibfnamefont {J.~P.}\ \bibnamefont {Eastwood}}, \bibinfo {author} {\bibfnamefont {T.}~\bibnamefont {Phan}}, \bibinfo {author} {\bibfnamefont {I.}~\bibnamefont {Gingell}}, \bibinfo {author} {\bibfnamefont {M.}~\bibnamefont {Shay}}, \bibinfo {author} {\bibfnamefont {J.}~\bibnamefont {Burch}}, \bibinfo {author} {\bibfnamefont {R.}~\bibnamefont {Ergun}}, \bibinfo {author} {\bibfnamefont {B.}~\bibnamefont {Giles}}, \bibinfo {author} {\bibfnamefont {D.}~\bibnamefont {Gershman}}, \bibinfo {author} {\bibfnamefont {O.}~\bibnamefont {Le~Contel}}, \emph {et~al.},\ }\bibfield  {title} {\bibinfo {title} {Properties of the turbulence associated with electron-only magnetic reconnection in earth’s magnetosheath},\ }\href@noop {} {\bibfield  {journal} {\bibinfo  {journal} {The Astrophysical journal letters}\ }\textbf {\bibinfo {volume} {877}},\ \bibinfo {pages} {L37} (\bibinfo {year} {2019})}\BibitemShut {NoStop}%
\bibitem [{\citenamefont {Jain}\ \emph {et~al.}(2012)\citenamefont {Jain}, \citenamefont {Sharma}, \citenamefont {Zelenyi},\ and\ \citenamefont {Malova}}]{jain2012electron}%
  \BibitemOpen
  \bibfield  {author} {\bibinfo {author} {\bibfnamefont {N.}~\bibnamefont {Jain}}, \bibinfo {author} {\bibfnamefont {A.~S.}\ \bibnamefont {Sharma}}, \bibinfo {author} {\bibfnamefont {L.}~\bibnamefont {Zelenyi}},\ and\ \bibinfo {author} {\bibfnamefont {H.}~\bibnamefont {Malova}},\ }\bibfield  {title} {\bibinfo {title} {Electron scale structures of thin current sheets in magnetic reconnection},\ }in\ \href@noop {} {\emph {\bibinfo {booktitle} {Annales geophysicae}}},\ Vol.~\bibinfo {volume} {30}\ (\bibinfo {organization} {Copernicus Publications G{\"o}ttingen, Germany},\ \bibinfo {year} {2012})\ pp.\ \bibinfo {pages} {661--666}\BibitemShut {NoStop}%
\bibitem [{\citenamefont {Sharma~Pyakurel}\ \emph {et~al.}(2019)\citenamefont {Sharma~Pyakurel}, \citenamefont {Shay}, \citenamefont {Phan}, \citenamefont {Matthaeus}, \citenamefont {Drake}, \citenamefont {TenBarge}, \citenamefont {Haggerty}, \citenamefont {Klein}, \citenamefont {Cassak}, \citenamefont {Parashar} \emph {et~al.}}]{sharma2019transition}%
  \BibitemOpen
  \bibfield  {author} {\bibinfo {author} {\bibfnamefont {P.}~\bibnamefont {Sharma~Pyakurel}}, \bibinfo {author} {\bibfnamefont {M.}~\bibnamefont {Shay}}, \bibinfo {author} {\bibfnamefont {T.}~\bibnamefont {Phan}}, \bibinfo {author} {\bibfnamefont {W.}~\bibnamefont {Matthaeus}}, \bibinfo {author} {\bibfnamefont {J.}~\bibnamefont {Drake}}, \bibinfo {author} {\bibfnamefont {J.}~\bibnamefont {TenBarge}}, \bibinfo {author} {\bibfnamefont {C.}~\bibnamefont {Haggerty}}, \bibinfo {author} {\bibfnamefont {K.}~\bibnamefont {Klein}}, \bibinfo {author} {\bibfnamefont {P.}~\bibnamefont {Cassak}}, \bibinfo {author} {\bibfnamefont {T.}~\bibnamefont {Parashar}}, \emph {et~al.},\ }\bibfield  {title} {\bibinfo {title} {Transition from ion-coupled to electron-only reconnection: Basic physics and implications for plasma turbulence},\ }\href@noop {} {\bibfield  {journal} {\bibinfo  {journal} {Physics of Plasmas}\ }\textbf {\bibinfo {volume} {26}} (\bibinfo {year} {2019})}\BibitemShut {NoStop}%
\bibitem [{\citenamefont {Califano}\ \emph {et~al.}(2020)\citenamefont {Califano}, \citenamefont {Cerri}, \citenamefont {Faganello}, \citenamefont {Laveder}, \citenamefont {Sisti},\ and\ \citenamefont {Kunz}}]{califano2020electron}%
  \BibitemOpen
  \bibfield  {author} {\bibinfo {author} {\bibfnamefont {F.}~\bibnamefont {Califano}}, \bibinfo {author} {\bibfnamefont {S.~S.}\ \bibnamefont {Cerri}}, \bibinfo {author} {\bibfnamefont {M.}~\bibnamefont {Faganello}}, \bibinfo {author} {\bibfnamefont {D.}~\bibnamefont {Laveder}}, \bibinfo {author} {\bibfnamefont {M.}~\bibnamefont {Sisti}},\ and\ \bibinfo {author} {\bibfnamefont {M.~W.}\ \bibnamefont {Kunz}},\ }\bibfield  {title} {\bibinfo {title} {Electron-only reconnection in plasma turbulence},\ }\href@noop {} {\bibfield  {journal} {\bibinfo  {journal} {Frontiers in Physics}\ }\textbf {\bibinfo {volume} {8}},\ \bibinfo {pages} {317} (\bibinfo {year} {2020})}\BibitemShut {NoStop}%
\bibitem [{\citenamefont {Arr{\`o}}\ \emph {et~al.}(2020)\citenamefont {Arr{\`o}}, \citenamefont {Califano},\ and\ \citenamefont {Lapenta}}]{arro2020statistical}%
  \BibitemOpen
  \bibfield  {author} {\bibinfo {author} {\bibfnamefont {G.}~\bibnamefont {Arr{\`o}}}, \bibinfo {author} {\bibfnamefont {F.}~\bibnamefont {Califano}},\ and\ \bibinfo {author} {\bibfnamefont {G.}~\bibnamefont {Lapenta}},\ }\bibfield  {title} {\bibinfo {title} {Statistical properties of turbulent fluctuations associated with electron-only magnetic reconnection},\ }\href@noop {} {\bibfield  {journal} {\bibinfo  {journal} {Astronomy \& Astrophysics}\ }\textbf {\bibinfo {volume} {642}},\ \bibinfo {pages} {A45} (\bibinfo {year} {2020})}\BibitemShut {NoStop}%
\bibitem [{\citenamefont {Vega}\ \emph {et~al.}(2020)\citenamefont {Vega}, \citenamefont {Roytershteyn}, \citenamefont {Delzanno},\ and\ \citenamefont {Boldyrev}}]{vega2020electron}%
  \BibitemOpen
  \bibfield  {author} {\bibinfo {author} {\bibfnamefont {C.}~\bibnamefont {Vega}}, \bibinfo {author} {\bibfnamefont {V.}~\bibnamefont {Roytershteyn}}, \bibinfo {author} {\bibfnamefont {G.~L.}\ \bibnamefont {Delzanno}},\ and\ \bibinfo {author} {\bibfnamefont {S.}~\bibnamefont {Boldyrev}},\ }\bibfield  {title} {\bibinfo {title} {Electron-only reconnection in kinetic-{A}lfv{\'e}n turbulence},\ }\href@noop {} {\bibfield  {journal} {\bibinfo  {journal} {The Astrophysical Journal Letters}\ }\textbf {\bibinfo {volume} {893}},\ \bibinfo {pages} {L10} (\bibinfo {year} {2020})}\BibitemShut {NoStop}%
\bibitem [{\citenamefont {Greess}\ \emph {et~al.}(2022)\citenamefont {Greess}, \citenamefont {Egedal}, \citenamefont {Stanier}, \citenamefont {Olson}, \citenamefont {Daughton}, \citenamefont {L{\^e}}, \citenamefont {Millet-Ayala}, \citenamefont {Kuchta},\ and\ \citenamefont {Forest}}]{greess2022kinetic}%
  \BibitemOpen
  \bibfield  {author} {\bibinfo {author} {\bibfnamefont {S.}~\bibnamefont {Greess}}, \bibinfo {author} {\bibfnamefont {J.}~\bibnamefont {Egedal}}, \bibinfo {author} {\bibfnamefont {A.}~\bibnamefont {Stanier}}, \bibinfo {author} {\bibfnamefont {J.}~\bibnamefont {Olson}}, \bibinfo {author} {\bibfnamefont {W.}~\bibnamefont {Daughton}}, \bibinfo {author} {\bibfnamefont {A.}~\bibnamefont {L{\^e}}}, \bibinfo {author} {\bibfnamefont {A.}~\bibnamefont {Millet-Ayala}}, \bibinfo {author} {\bibfnamefont {C.}~\bibnamefont {Kuchta}},\ and\ \bibinfo {author} {\bibfnamefont {C.}~\bibnamefont {Forest}},\ }\bibfield  {title} {\bibinfo {title} {Kinetic simulations verifying reconnection rates measured in the laboratory, spanning the ion-coupled to near electron-only regimes},\ }\href@noop {} {\bibfield  {journal} {\bibinfo  {journal} {Physics of Plasmas}\ }\textbf {\bibinfo {volume} {29}} (\bibinfo {year} {2022})}\BibitemShut {NoStop}%
\bibitem [{\citenamefont {Guan}\ \emph {et~al.}(2023)\citenamefont {Guan}, \citenamefont {Lu}, \citenamefont {Lu}, \citenamefont {Huang},\ and\ \citenamefont {Wang}}]{guan2023reconnection}%
  \BibitemOpen
  \bibfield  {author} {\bibinfo {author} {\bibfnamefont {Y.}~\bibnamefont {Guan}}, \bibinfo {author} {\bibfnamefont {Q.}~\bibnamefont {Lu}}, \bibinfo {author} {\bibfnamefont {S.}~\bibnamefont {Lu}}, \bibinfo {author} {\bibfnamefont {K.}~\bibnamefont {Huang}},\ and\ \bibinfo {author} {\bibfnamefont {R.}~\bibnamefont {Wang}},\ }\bibfield  {title} {\bibinfo {title} {Reconnection rate and transition from ion-coupled to electron-only reconnection},\ }\href@noop {} {\bibfield  {journal} {\bibinfo  {journal} {The Astrophysical Journal}\ }\textbf {\bibinfo {volume} {958}},\ \bibinfo {pages} {172} (\bibinfo {year} {2023})}\BibitemShut {NoStop}%
\bibitem [{\citenamefont {{Schekochihin}}\ \emph {et~al.}(2009)\citenamefont {{Schekochihin}}, \citenamefont {{Cowley}}, \citenamefont {{Dorland}}, \citenamefont {{Hammett}}, \citenamefont {{Howes}}, \citenamefont {{Quataert}},\ and\ \citenamefont {{Tatsuno}}}]{schekochihin2009astrophysical}%
  \BibitemOpen
  \bibfield  {author} {\bibinfo {author} {\bibfnamefont {A.~A.}\ \bibnamefont {{Schekochihin}}}, \bibinfo {author} {\bibfnamefont {S.~C.}\ \bibnamefont {{Cowley}}}, \bibinfo {author} {\bibfnamefont {W.}~\bibnamefont {{Dorland}}}, \bibinfo {author} {\bibfnamefont {G.~W.}\ \bibnamefont {{Hammett}}}, \bibinfo {author} {\bibfnamefont {G.~G.}\ \bibnamefont {{Howes}}}, \bibinfo {author} {\bibfnamefont {E.}~\bibnamefont {{Quataert}}},\ and\ \bibinfo {author} {\bibfnamefont {T.}~\bibnamefont {{Tatsuno}}},\ }\bibfield  {title} {\bibinfo {title} {Astrophysical gyrokinetics: kinetic and fluid turbulent cascades in magnetized weakly collisional plasmas},\ }\href@noop {} {\bibfield  {journal} {\bibinfo  {journal} {The Astrophysical Journal Supplement Series}\ }\textbf {\bibinfo {volume} {182}},\ \bibinfo {pages} {310} (\bibinfo {year} {2009})}\BibitemShut {NoStop}%
\bibitem [{\citenamefont {Zocco}\ and\ \citenamefont {Schekochihin}(2011)}]{zocco2011reduced}%
  \BibitemOpen
  \bibfield  {author} {\bibinfo {author} {\bibfnamefont {A.}~\bibnamefont {Zocco}}\ and\ \bibinfo {author} {\bibfnamefont {A.~A.}\ \bibnamefont {Schekochihin}},\ }\bibfield  {title} {\bibinfo {title} {Reduced fluid-kinetic equations for low-frequency dynamics, magnetic reconnection, and electron heating in low-beta plasmas},\ }\href@noop {} {\bibfield  {journal} {\bibinfo  {journal} {Physics of Plasmas}\ }\textbf {\bibinfo {volume} {18}} (\bibinfo {year} {2011})}\BibitemShut {NoStop}%
\bibitem [{\citenamefont {Passot}\ \emph {et~al.}(2018)\citenamefont {Passot}, \citenamefont {Sulem},\ and\ \citenamefont {Tassi}}]{passot2018gyrofluid}%
  \BibitemOpen
  \bibfield  {author} {\bibinfo {author} {\bibfnamefont {T.}~\bibnamefont {Passot}}, \bibinfo {author} {\bibfnamefont {P.~L.}\ \bibnamefont {Sulem}},\ and\ \bibinfo {author} {\bibfnamefont {E.}~\bibnamefont {Tassi}},\ }\bibfield  {title} {\bibinfo {title} {Gyrofluid modeling and phenomenology of low-$\beta$e {A}lfv{\'e}n wave turbulence},\ }\href@noop {} {\bibfield  {journal} {\bibinfo  {journal} {Physics of Plasmas}\ }\textbf {\bibinfo {volume} {25}},\ \bibinfo {pages} {042107} (\bibinfo {year} {2018})}\BibitemShut {NoStop}%
\bibitem [{\citenamefont {Meyrand}\ \emph {et~al.}(2021)\citenamefont {Meyrand}, \citenamefont {Squire}, \citenamefont {Schekochihin},\ and\ \citenamefont {Dorland}}]{meyrand2021violation}%
  \BibitemOpen
  \bibfield  {author} {\bibinfo {author} {\bibfnamefont {R.}~\bibnamefont {Meyrand}}, \bibinfo {author} {\bibfnamefont {J.}~\bibnamefont {Squire}}, \bibinfo {author} {\bibfnamefont {A.~A.}\ \bibnamefont {Schekochihin}},\ and\ \bibinfo {author} {\bibfnamefont {W.}~\bibnamefont {Dorland}},\ }\bibfield  {title} {\bibinfo {title} {On the violation of the zeroth law of turbulence in space plasmas},\ }\href@noop {} {\bibfield  {journal} {\bibinfo  {journal} {Journal of Plasma Physics}\ }\textbf {\bibinfo {volume} {87}} (\bibinfo {year} {2021})}\BibitemShut {NoStop}%
\bibitem [{\citenamefont {Zhou}\ \emph {et~al.}(2023{\natexlab{a}})\citenamefont {Zhou}, \citenamefont {Liu},\ and\ \citenamefont {Loureiro}}]{zhou2023spectrum}%
  \BibitemOpen
  \bibfield  {author} {\bibinfo {author} {\bibfnamefont {M.}~\bibnamefont {Zhou}}, \bibinfo {author} {\bibfnamefont {Z.}~\bibnamefont {Liu}},\ and\ \bibinfo {author} {\bibfnamefont {N.~F.}\ \bibnamefont {Loureiro}},\ }\bibfield  {title} {\bibinfo {title} {Spectrum of kinetic-{A}lfv{\'e}n-wave turbulence: intermittency or tearing mediation?},\ }\href@noop {} {\bibfield  {journal} {\bibinfo  {journal} {Monthly Notices of the Royal Astronomical Society}\ }\textbf {\bibinfo {volume} {524}},\ \bibinfo {pages} {5468} (\bibinfo {year} {2023}{\natexlab{a}})}\BibitemShut {NoStop}%
\bibitem [{Note1()}]{Note1}%
  \BibitemOpen
  \bibinfo {note} {These equations are the isothermal limit of the Kinetic Reduced Electron Heating Model (KREHM), a rigorous analytical framework derived from gyrokinetics under the condition of low electron plasma beta ($\beta _e \sim m_e/m_i$, where $\beta _e$ is the ratio of the electron to the magnetic pressure, and $m_e$ and $m_i$ are the electron and proton masses)~\protect \citep {zocco2011reduced}. See Supplemental Material for more details and results in the non-isothermal limit.}\BibitemShut {Stop}%
\bibitem [{\citenamefont {Krommes}(2002)}]{krommes2002fundamental}%
  \BibitemOpen
  \bibfield  {author} {\bibinfo {author} {\bibfnamefont {J.~A.}\ \bibnamefont {Krommes}},\ }\bibfield  {title} {\bibinfo {title} {Fundamental statistical descriptions of plasma turbulence in magnetic fields},\ }\href@noop {} {\bibfield  {journal} {\bibinfo  {journal} {Physics Reports}\ }\textbf {\bibinfo {volume} {360}},\ \bibinfo {pages} {1} (\bibinfo {year} {2002})}\BibitemShut {NoStop}%
\bibitem [{Note38()}]{Note38}%
  \BibitemOpen
  \bibinfo {note} {We use $<$ instead of $\ll $ in order not to interfere with the ordering under which Eqs.~(\ref {eq:ne}-\ref {eq:gk_poisson}) are derived (namely, $\beta \sim m_e/m_i$ so that $\rho _i \sim d_e$, see~\protect \citep {zocco2011reduced}). We note that $\beta $ does not have to be numerically close to $m_e/m_i$; all that is formally required is for $\rho _i/d_e$ not to be so large as to interfere with the asymptotic expansion parameters. Empirically, it is known that Eqs.~(\ref {eq:ne}-\ref {eq:gk_poisson}) remain valid at least up to $\beta \sim 0.1$ (see ~\protect \citep {grovselj2017fully}). In practice, a maximum ratio of $\rho _i/d_e\approx 20$ is acceptable if adopting the real proton-electron mass ratio.}\BibitemShut {Stop}%
\bibitem [{\citenamefont {Birn}\ \emph {et~al.}(2001)\citenamefont {Birn}, \citenamefont {Drake}, \citenamefont {Shay}, \citenamefont {Rogers}, \citenamefont {Denton}, \citenamefont {Hesse}, \citenamefont {Kuznetsova}, \citenamefont {Ma}, \citenamefont {Bhattacharjee}, \citenamefont {Otto},\ and\ \citenamefont {Pritchett}}]{birn2001GEM}%
  \BibitemOpen
  \bibfield  {author} {\bibinfo {author} {\bibfnamefont {J.}~\bibnamefont {Birn}}, \bibinfo {author} {\bibfnamefont {J.~F.}\ \bibnamefont {Drake}}, \bibinfo {author} {\bibfnamefont {M.~A.}\ \bibnamefont {Shay}}, \bibinfo {author} {\bibfnamefont {B.~N.}\ \bibnamefont {Rogers}}, \bibinfo {author} {\bibfnamefont {R.~E.}\ \bibnamefont {Denton}}, \bibinfo {author} {\bibfnamefont {M.}~\bibnamefont {Hesse}}, \bibinfo {author} {\bibfnamefont {M.}~\bibnamefont {Kuznetsova}}, \bibinfo {author} {\bibfnamefont {Z.~W.}\ \bibnamefont {Ma}}, \bibinfo {author} {\bibfnamefont {A.}~\bibnamefont {Bhattacharjee}}, \bibinfo {author} {\bibfnamefont {A.}~\bibnamefont {Otto}},\ and\ \bibinfo {author} {\bibfnamefont {P.~L.}\ \bibnamefont {Pritchett}},\ }\bibfield  {title} {\bibinfo {title} {Geospace environmental modeling (gem) magnetic reconnection challenge},\ }\href {https://doi.org/https://doi.org/10.1029/1999JA900449} {\bibfield  {journal} {\bibinfo  {journal} {Journal of Geophysical Research: Space Physics}\ }\textbf {\bibinfo
  {volume} {106}},\ \bibinfo {pages} {3715} (\bibinfo {year} {2001})}\BibitemShut {NoStop}%
\bibitem [{\citenamefont {Cassak}\ \emph {et~al.}(2017)\citenamefont {Cassak}, \citenamefont {Liu},\ and\ \citenamefont {Shay}}]{cassak2017review}%
  \BibitemOpen
  \bibfield  {author} {\bibinfo {author} {\bibfnamefont {P.}~\bibnamefont {Cassak}}, \bibinfo {author} {\bibfnamefont {Y.-H.}\ \bibnamefont {Liu}},\ and\ \bibinfo {author} {\bibfnamefont {M.}~\bibnamefont {Shay}},\ }\bibfield  {title} {\bibinfo {title} {A review of the 0.1 reconnection rate problem},\ }\href@noop {} {\bibfield  {journal} {\bibinfo  {journal} {Journal of Plasma Physics}\ }\textbf {\bibinfo {volume} {83}},\ \bibinfo {pages} {715830501} (\bibinfo {year} {2017})}\BibitemShut {NoStop}%
\bibitem [{\citenamefont {Liu}\ \emph {et~al.}(2017)\citenamefont {Liu}, \citenamefont {Hesse}, \citenamefont {Guo}, \citenamefont {Daughton}, \citenamefont {Li}, \citenamefont {Cassak},\ and\ \citenamefont {Shay}}]{liu2017does}%
  \BibitemOpen
  \bibfield  {author} {\bibinfo {author} {\bibfnamefont {Y.-H.}\ \bibnamefont {Liu}}, \bibinfo {author} {\bibfnamefont {M.}~\bibnamefont {Hesse}}, \bibinfo {author} {\bibfnamefont {F.}~\bibnamefont {Guo}}, \bibinfo {author} {\bibfnamefont {W.}~\bibnamefont {Daughton}}, \bibinfo {author} {\bibfnamefont {H.}~\bibnamefont {Li}}, \bibinfo {author} {\bibfnamefont {P.}~\bibnamefont {Cassak}},\ and\ \bibinfo {author} {\bibfnamefont {M.}~\bibnamefont {Shay}},\ }\bibfield  {title} {\bibinfo {title} {Why does steady-state magnetic reconnection have a maximum local rate of order 0.1?},\ }\href@noop {} {\bibfield  {journal} {\bibinfo  {journal} {Physical Review Letters}\ }\textbf {\bibinfo {volume} {118}},\ \bibinfo {pages} {085101} (\bibinfo {year} {2017})}\BibitemShut {NoStop}%
\bibitem [{\citenamefont {Shi}\ \emph {et~al.}(2022)\citenamefont {Shi}, \citenamefont {Srivastav}, \citenamefont {Barbhuiya}, \citenamefont {Cassak}, \citenamefont {Scime},\ and\ \citenamefont {Swisdak}}]{shi2022laboratory}%
  \BibitemOpen
  \bibfield  {author} {\bibinfo {author} {\bibfnamefont {P.}~\bibnamefont {Shi}}, \bibinfo {author} {\bibfnamefont {P.}~\bibnamefont {Srivastav}}, \bibinfo {author} {\bibfnamefont {M.~H.}\ \bibnamefont {Barbhuiya}}, \bibinfo {author} {\bibfnamefont {P.~A.}\ \bibnamefont {Cassak}}, \bibinfo {author} {\bibfnamefont {E.~E.}\ \bibnamefont {Scime}},\ and\ \bibinfo {author} {\bibfnamefont {M.}~\bibnamefont {Swisdak}},\ }\bibfield  {title} {\bibinfo {title} {Laboratory observations of electron heating and non-maxwellian distributions at the kinetic scale during electron-only magnetic reconnection},\ }\href@noop {} {\bibfield  {journal} {\bibinfo  {journal} {Physical Review Letters}\ }\textbf {\bibinfo {volume} {128}},\ \bibinfo {pages} {025002} (\bibinfo {year} {2022})}\BibitemShut {NoStop}%
\bibitem [{\citenamefont {Mandell}\ \emph {et~al.}(2018)\citenamefont {Mandell}, \citenamefont {Dorland},\ and\ \citenamefont {Landreman}}]{mandell2018laguerre}%
  \BibitemOpen
  \bibfield  {author} {\bibinfo {author} {\bibfnamefont {N.}~\bibnamefont {Mandell}}, \bibinfo {author} {\bibfnamefont {W.}~\bibnamefont {Dorland}},\ and\ \bibinfo {author} {\bibfnamefont {M.}~\bibnamefont {Landreman}},\ }\bibfield  {title} {\bibinfo {title} {Laguerre--{H}ermite pseudo-spectral velocity formulation of gyrokinetics},\ }\href@noop {} {\bibfield  {journal} {\bibinfo  {journal} {Journal of Plasma Physics}\ }\textbf {\bibinfo {volume} {84}},\ \bibinfo {pages} {905840108} (\bibinfo {year} {2018})}\BibitemShut {NoStop}%
\bibitem [{\citenamefont {Mandell}\ \emph {et~al.}(2022)\citenamefont {Mandell}, \citenamefont {Dorland}, \citenamefont {Abel}, \citenamefont {Gaur}, \citenamefont {Kim}, \citenamefont {Martin},\ and\ \citenamefont {Qian}}]{mandell2022gx}%
  \BibitemOpen
  \bibfield  {author} {\bibinfo {author} {\bibfnamefont {N.}~\bibnamefont {Mandell}}, \bibinfo {author} {\bibfnamefont {W.}~\bibnamefont {Dorland}}, \bibinfo {author} {\bibfnamefont {I.}~\bibnamefont {Abel}}, \bibinfo {author} {\bibfnamefont {R.}~\bibnamefont {Gaur}}, \bibinfo {author} {\bibfnamefont {P.}~\bibnamefont {Kim}}, \bibinfo {author} {\bibfnamefont {M.}~\bibnamefont {Martin}},\ and\ \bibinfo {author} {\bibfnamefont {T.}~\bibnamefont {Qian}},\ }\bibfield  {title} {\bibinfo {title} {Gx: a gpu-native gyrokinetic turbulence code for tokamak and stellarator design},\ }\href@noop {} {\bibfield  {journal} {\bibinfo  {journal} {arXiv preprint arXiv:2209.06731}\ } (\bibinfo {year} {2022})}\BibitemShut {NoStop}%
\bibitem [{\citenamefont {Rogers}\ and\ \citenamefont {Zakharov}(1996)}]{rogers1996collisionless}%
  \BibitemOpen
  \bibfield  {author} {\bibinfo {author} {\bibfnamefont {B.}~\bibnamefont {Rogers}}\ and\ \bibinfo {author} {\bibfnamefont {L.}~\bibnamefont {Zakharov}},\ }\bibfield  {title} {\bibinfo {title} {Collisionless m= 1 reconnection in tokamaks},\ }\href@noop {} {\bibfield  {journal} {\bibinfo  {journal} {Physics of Plasmas}\ }\textbf {\bibinfo {volume} {3}},\ \bibinfo {pages} {2411} (\bibinfo {year} {1996})}\BibitemShut {NoStop}%
\bibitem [{\citenamefont {Grasso}\ \emph {et~al.}(2000)\citenamefont {Grasso}, \citenamefont {Califano}, \citenamefont {Pegoraro},\ and\ \citenamefont {Porcelli}}]{grasso2000ion}%
  \BibitemOpen
  \bibfield  {author} {\bibinfo {author} {\bibfnamefont {D.}~\bibnamefont {Grasso}}, \bibinfo {author} {\bibfnamefont {F.}~\bibnamefont {Califano}}, \bibinfo {author} {\bibfnamefont {F.}~\bibnamefont {Pegoraro}},\ and\ \bibinfo {author} {\bibfnamefont {F.}~\bibnamefont {Porcelli}},\ }\bibfield  {title} {\bibinfo {title} {Ion larmor radius effects in collisionless reconnection},\ }\href@noop {} {\bibfield  {journal} {\bibinfo  {journal} {Plasma Physics Reports}\ }\textbf {\bibinfo {volume} {26}},\ \bibinfo {pages} {512} (\bibinfo {year} {2000})}\BibitemShut {NoStop}%
\bibitem [{Note2()}]{Note2}%
  \BibitemOpen
  \bibinfo {note} {This initial configuration is approximately an equilibrium but it is unstable to coalescence instability~\cite {finn1976coalescence}. We choose this configuration instead of the typical Harris sheet~\cite {harris1962plasma} for consistency with the study of the inverse magnetic energy transfer problem discussed below. However, we do not expect the results to be sensitive to the specific configuration chosen.}\BibitemShut {Stop}%
\bibitem [{\citenamefont {Isobe}\ \emph {et~al.}(2005)\citenamefont {Isobe}, \citenamefont {Takasaki},\ and\ \citenamefont {Shibata}}]{Isobe2005MeasurementOT}%
  \BibitemOpen
  \bibfield  {author} {\bibinfo {author} {\bibfnamefont {H.}~\bibnamefont {Isobe}}, \bibinfo {author} {\bibfnamefont {H.}~\bibnamefont {Takasaki}},\ and\ \bibinfo {author} {\bibfnamefont {K.}~\bibnamefont {Shibata}},\ }\bibfield  {title} {\bibinfo {title} {Measurement of the energy release rate and the reconnection rate in solar flares},\ }\href@noop {} {\bibfield  {journal} {\bibinfo  {journal} {The Astrophysical Journal}\ }\textbf {\bibinfo {volume} {632}},\ \bibinfo {pages} {1184} (\bibinfo {year} {2005})}\BibitemShut {NoStop}%
\bibitem [{\citenamefont {Zenitani}\ and\ \citenamefont {Hesse}(2008)}]{Zenitani2008}%
  \BibitemOpen
  \bibfield  {author} {\bibinfo {author} {\bibfnamefont {S.}~\bibnamefont {Zenitani}}\ and\ \bibinfo {author} {\bibfnamefont {M.}~\bibnamefont {Hesse}},\ }\bibfield  {title} {\bibinfo {title} {The role of the {W}eibel instability at the reconnection jet front in relativistic pair plasma reconnection},\ }\href {https://doi.org/10.1063/1.2836623} {\bibfield  {journal} {\bibinfo  {journal} {Physics of Plasmas}\ }\textbf {\bibinfo {volume} {15}},\ \bibinfo {pages} {022101} (\bibinfo {year} {2008})},\ \Eprint {https://arxiv.org/abs/https://doi.org/10.1063/1.2836623} {https://doi.org/10.1063/1.2836623} \BibitemShut {NoStop}%
\bibitem [{\citenamefont {Drake}\ \emph {et~al.}(2008)\citenamefont {Drake}, \citenamefont {Shay},\ and\ \citenamefont {Swisdak}}]{Drake2008}%
  \BibitemOpen
  \bibfield  {author} {\bibinfo {author} {\bibfnamefont {J.~F.}\ \bibnamefont {Drake}}, \bibinfo {author} {\bibfnamefont {M.~A.}\ \bibnamefont {Shay}},\ and\ \bibinfo {author} {\bibfnamefont {M.}~\bibnamefont {Swisdak}},\ }\bibfield  {title} {\bibinfo {title} {The {H}all fields and fast magnetic reconnection},\ }\href {https://doi.org/10.1063/1.2901194} {\bibfield  {journal} {\bibinfo  {journal} {Physics of Plasmas}\ }\textbf {\bibinfo {volume} {15}},\ \bibinfo {pages} {042306} (\bibinfo {year} {2008})},\ \Eprint {https://arxiv.org/abs/https://aip.scitation.org/doi/pdf/10.1063/1.2901194} {https://aip.scitation.org/doi/pdf/10.1063/1.2901194} \BibitemShut {NoStop}%
\bibitem [{\citenamefont {Bessho}\ and\ \citenamefont {Bhattacharjee}(2012)}]{Bessho_2012}%
  \BibitemOpen
  \bibfield  {author} {\bibinfo {author} {\bibfnamefont {N.}~\bibnamefont {Bessho}}\ and\ \bibinfo {author} {\bibfnamefont {A.}~\bibnamefont {Bhattacharjee}},\ }\bibfield  {title} {\bibinfo {title} {Fast magnetic reconnection and particle acceleration in relativistic low-density electron--positron plasmas without guide field},\ }\href@noop {} {\bibfield  {journal} {\bibinfo  {journal} {The Astrophysical Journal}\ }\textbf {\bibinfo {volume} {750}},\ \bibinfo {pages} {129} (\bibinfo {year} {2012})}\BibitemShut {NoStop}%
\bibitem [{\citenamefont {Chen}\ \emph {et~al.}(2017)\citenamefont {Chen}, \citenamefont {Hesse}, \citenamefont {Wang}, \citenamefont {Gershman}, \citenamefont {Ergun}, \citenamefont {Burch}, \citenamefont {Bessho}, \citenamefont {Torbert}, \citenamefont {Giles}, \citenamefont {Webster} \emph {et~al.}}]{Chen2017}%
  \BibitemOpen
  \bibfield  {author} {\bibinfo {author} {\bibfnamefont {L.-J.}\ \bibnamefont {Chen}}, \bibinfo {author} {\bibfnamefont {M.}~\bibnamefont {Hesse}}, \bibinfo {author} {\bibfnamefont {S.}~\bibnamefont {Wang}}, \bibinfo {author} {\bibfnamefont {D.}~\bibnamefont {Gershman}}, \bibinfo {author} {\bibfnamefont {R.}~\bibnamefont {Ergun}}, \bibinfo {author} {\bibfnamefont {J.}~\bibnamefont {Burch}}, \bibinfo {author} {\bibfnamefont {N.}~\bibnamefont {Bessho}}, \bibinfo {author} {\bibfnamefont {R.}~\bibnamefont {Torbert}}, \bibinfo {author} {\bibfnamefont {B.}~\bibnamefont {Giles}}, \bibinfo {author} {\bibfnamefont {J.}~\bibnamefont {Webster}}, \emph {et~al.},\ }\bibfield  {title} {\bibinfo {title} {Electron diffusion region during magnetopause reconnection with an intermediate guide field: Magnetospheric multiscale observations},\ }\href@noop {} {\bibfield  {journal} {\bibinfo  {journal} {Journal of Geophysical Research: Space Physics}\ }\textbf {\bibinfo {volume} {122}},\ \bibinfo {pages} {5235} (\bibinfo {year}
  {2017})}\BibitemShut {NoStop}%
\bibitem [{\citenamefont {Loureiro}\ \emph {et~al.}(2013)\citenamefont {Loureiro}, \citenamefont {Schekochihin},\ and\ \citenamefont {Zocco}}]{loureiro2013fast}%
  \BibitemOpen
  \bibfield  {author} {\bibinfo {author} {\bibfnamefont {N.}~\bibnamefont {Loureiro}}, \bibinfo {author} {\bibfnamefont {A.}~\bibnamefont {Schekochihin}},\ and\ \bibinfo {author} {\bibfnamefont {A.}~\bibnamefont {Zocco}},\ }\bibfield  {title} {\bibinfo {title} {Fast collisionless reconnection and electron heating in strongly magnetized plasmas},\ }\href@noop {} {\bibfield  {journal} {\bibinfo  {journal} {Physical Review Letters}\ }\textbf {\bibinfo {volume} {111}},\ \bibinfo {pages} {025002} (\bibinfo {year} {2013})}\BibitemShut {NoStop}%
\bibitem [{\citenamefont {Numata}\ and\ \citenamefont {Loureiro}(2015)}]{numata2015ion}%
  \BibitemOpen
  \bibfield  {author} {\bibinfo {author} {\bibfnamefont {R.}~\bibnamefont {Numata}}\ and\ \bibinfo {author} {\bibfnamefont {N.}~\bibnamefont {Loureiro}},\ }\bibfield  {title} {\bibinfo {title} {Ion and electron heating during magnetic reconnection in weakly collisional plasmas},\ }\href@noop {} {\bibfield  {journal} {\bibinfo  {journal} {Journal of Plasma Physics}\ }\textbf {\bibinfo {volume} {81}},\ \bibinfo {pages} {305810201} (\bibinfo {year} {2015})}\BibitemShut {NoStop}%
\bibitem [{\citenamefont {Mallet}(2020)}]{mallet2020onset}%
  \BibitemOpen
  \bibfield  {author} {\bibinfo {author} {\bibfnamefont {A.}~\bibnamefont {Mallet}},\ }\bibfield  {title} {\bibinfo {title} {The onset of electron-only reconnection},\ }\href@noop {} {\bibfield  {journal} {\bibinfo  {journal} {Journal of Plasma Physics}\ }\textbf {\bibinfo {volume} {86}},\ \bibinfo {pages} {905860301} (\bibinfo {year} {2020})}\BibitemShut {NoStop}%
\bibitem [{\citenamefont {Goldstein}\ \emph {et~al.}(1995)\citenamefont {Goldstein}, \citenamefont {Smith}, \citenamefont {Balogh}, \citenamefont {Horbury}, \citenamefont {Goldstein},\ and\ \citenamefont {Roberts}}]{goldstein1995properties}%
  \BibitemOpen
  \bibfield  {author} {\bibinfo {author} {\bibfnamefont {B.}~\bibnamefont {Goldstein}}, \bibinfo {author} {\bibfnamefont {E.}~\bibnamefont {Smith}}, \bibinfo {author} {\bibfnamefont {A.}~\bibnamefont {Balogh}}, \bibinfo {author} {\bibfnamefont {T.}~\bibnamefont {Horbury}}, \bibinfo {author} {\bibfnamefont {M.}~\bibnamefont {Goldstein}},\ and\ \bibinfo {author} {\bibfnamefont {D.}~\bibnamefont {Roberts}},\ }\bibfield  {title} {\bibinfo {title} {Properties of magnetohydrodynamic turbulence in the solar wind as observed by ulysses at high heliographic latitudes},\ }\href@noop {} {\bibfield  {journal} {\bibinfo  {journal} {Geophysical Research Letters}\ }\textbf {\bibinfo {volume} {22}},\ \bibinfo {pages} {3393} (\bibinfo {year} {1995})}\BibitemShut {NoStop}%
\bibitem [{Note3()}]{Note3}%
  \BibitemOpen
  \bibinfo {note} {The fit is slightly worse for $k_{\perp ,\protect \rm {max}}$ than for $\protect \mathcal {E}$ due to the discrete nature of this diagnostic and the limited change in the value of $k_{\perp ,\protect \rm {max}}$, given the constrained size of the simulation domain.}\BibitemShut {Stop}%
\bibitem [{\citenamefont {Cho}\ and\ \citenamefont {Lazarian}(2009)}]{cho2009simulations}%
  \BibitemOpen
  \bibfield  {author} {\bibinfo {author} {\bibfnamefont {J.}~\bibnamefont {Cho}}\ and\ \bibinfo {author} {\bibfnamefont {A.}~\bibnamefont {Lazarian}},\ }\bibfield  {title} {\bibinfo {title} {Simulations of electron magnetohydrodynamic turbulence},\ }\href@noop {} {\bibfield  {journal} {\bibinfo  {journal} {The Astrophysical Journal}\ }\textbf {\bibinfo {volume} {701}},\ \bibinfo {pages} {236} (\bibinfo {year} {2009})}\BibitemShut {NoStop}%
\bibitem [{\citenamefont {Cerri}\ \emph {et~al.}(2019)\citenamefont {Cerri}, \citenamefont {Gro{\v{s}}elj},\ and\ \citenamefont {Franci}}]{cerri2019kinetic}%
  \BibitemOpen
  \bibfield  {author} {\bibinfo {author} {\bibfnamefont {S.~S.}\ \bibnamefont {Cerri}}, \bibinfo {author} {\bibfnamefont {D.}~\bibnamefont {Gro{\v{s}}elj}},\ and\ \bibinfo {author} {\bibfnamefont {L.}~\bibnamefont {Franci}},\ }\bibfield  {title} {\bibinfo {title} {Kinetic plasma turbulence: recent insights and open questions from 3d3v simulations},\ }\href@noop {} {\bibfield  {journal} {\bibinfo  {journal} {Frontiers in Astronomy and Space Sciences}\ }\textbf {\bibinfo {volume} {6}},\ \bibinfo {pages} {64} (\bibinfo {year} {2019})}\BibitemShut {NoStop}%
\bibitem [{Note4()}]{Note4}%
  \BibitemOpen
  \bibinfo {note} {Before the system settles into a natural state of decaying turbulence described by our theory, it undergoes a transient stage characterized by the development of strong turbulence. An essential indicator of strong turbulence is the establishment of critical balance, where flux tubes initially confined to box length break in the $z$-direction. This reduction in the parallel coherence length $l$ is captured by the first two data points in Fig.~\ref {fig:2.3d} (b).}\BibitemShut {Stop}%
\bibitem [{\citenamefont {Boldyrev}\ and\ \citenamefont {Perez}(2012)}]{boldyrev2012spectrum}%
  \BibitemOpen
  \bibfield  {author} {\bibinfo {author} {\bibfnamefont {S.}~\bibnamefont {Boldyrev}}\ and\ \bibinfo {author} {\bibfnamefont {J.~C.}\ \bibnamefont {Perez}},\ }\bibfield  {title} {\bibinfo {title} {Spectrum of kinetic-{A}fv{\'e}n turbulence},\ }\href@noop {} {\bibfield  {journal} {\bibinfo  {journal} {The Astrophysical Journal Letters}\ }\textbf {\bibinfo {volume} {758}},\ \bibinfo {pages} {L44} (\bibinfo {year} {2012})}\BibitemShut {NoStop}%
\bibitem [{\citenamefont {Eastwood}\ \emph {et~al.}(2016)\citenamefont {Eastwood}, \citenamefont {Phan}, \citenamefont {Cassak}, \citenamefont {Gershman}, \citenamefont {Haggerty}, \citenamefont {Malakit}, \citenamefont {Shay}, \citenamefont {Mistry}, \citenamefont {{\O}ieroset}, \citenamefont {Russell} \emph {et~al.}}]{eastwood2016ion}%
  \BibitemOpen
  \bibfield  {author} {\bibinfo {author} {\bibfnamefont {J.}~\bibnamefont {Eastwood}}, \bibinfo {author} {\bibfnamefont {T.}~\bibnamefont {Phan}}, \bibinfo {author} {\bibfnamefont {P.}~\bibnamefont {Cassak}}, \bibinfo {author} {\bibfnamefont {D.}~\bibnamefont {Gershman}}, \bibinfo {author} {\bibfnamefont {C.}~\bibnamefont {Haggerty}}, \bibinfo {author} {\bibfnamefont {K.}~\bibnamefont {Malakit}}, \bibinfo {author} {\bibfnamefont {M.}~\bibnamefont {Shay}}, \bibinfo {author} {\bibfnamefont {R.}~\bibnamefont {Mistry}}, \bibinfo {author} {\bibfnamefont {M.}~\bibnamefont {{\O}ieroset}}, \bibinfo {author} {\bibfnamefont {C.}~\bibnamefont {Russell}}, \emph {et~al.},\ }\bibfield  {title} {\bibinfo {title} {Ion-scale secondary flux ropes generated by magnetopause reconnection as resolved by mms},\ }\href@noop {} {\bibfield  {journal} {\bibinfo  {journal} {Geophysical research letters}\ }\textbf {\bibinfo {volume} {43}},\ \bibinfo {pages} {4716} (\bibinfo {year} {2016})}\BibitemShut {NoStop}%
\bibitem [{\citenamefont {Sun}\ \emph {et~al.}(2019)\citenamefont {Sun}, \citenamefont {Slavin}, \citenamefont {Tian}, \citenamefont {Bai}, \citenamefont {Poh}, \citenamefont {Akhavan-Tafti}, \citenamefont {Lu}, \citenamefont {Yao}, \citenamefont {Le}, \citenamefont {Nakamura} \emph {et~al.}}]{sun2019mms}%
  \BibitemOpen
  \bibfield  {author} {\bibinfo {author} {\bibfnamefont {W.}~\bibnamefont {Sun}}, \bibinfo {author} {\bibfnamefont {J.}~\bibnamefont {Slavin}}, \bibinfo {author} {\bibfnamefont {A.}~\bibnamefont {Tian}}, \bibinfo {author} {\bibfnamefont {S.}~\bibnamefont {Bai}}, \bibinfo {author} {\bibfnamefont {G.}~\bibnamefont {Poh}}, \bibinfo {author} {\bibfnamefont {M.}~\bibnamefont {Akhavan-Tafti}}, \bibinfo {author} {\bibfnamefont {S.}~\bibnamefont {Lu}}, \bibinfo {author} {\bibfnamefont {S.}~\bibnamefont {Yao}}, \bibinfo {author} {\bibfnamefont {G.}~\bibnamefont {Le}}, \bibinfo {author} {\bibfnamefont {R.}~\bibnamefont {Nakamura}}, \emph {et~al.},\ }\bibfield  {title} {\bibinfo {title} {Mms study of the structure of ion-scale flux ropes in the earth's cross-tail current sheet},\ }\href@noop {} {\bibfield  {journal} {\bibinfo  {journal} {Geophysical Research Letters}\ }\textbf {\bibinfo {volume} {46}},\ \bibinfo {pages} {6168} (\bibinfo {year} {2019})}\BibitemShut {NoStop}%
\bibitem [{\citenamefont {Zhou}\ \emph {et~al.}(2022)\citenamefont {Zhou}, \citenamefont {Zhdankin}, \citenamefont {Kunz}, \citenamefont {Loureiro},\ and\ \citenamefont {Uzdensky}}]{zhou2022spontaneous}%
  \BibitemOpen
  \bibfield  {author} {\bibinfo {author} {\bibfnamefont {M.}~\bibnamefont {Zhou}}, \bibinfo {author} {\bibfnamefont {V.}~\bibnamefont {Zhdankin}}, \bibinfo {author} {\bibfnamefont {M.~W.}\ \bibnamefont {Kunz}}, \bibinfo {author} {\bibfnamefont {N.~F.}\ \bibnamefont {Loureiro}},\ and\ \bibinfo {author} {\bibfnamefont {D.~A.}\ \bibnamefont {Uzdensky}},\ }\bibfield  {title} {\bibinfo {title} {Spontaneous magnetization of collisionless plasma},\ }\href@noop {} {\bibfield  {journal} {\bibinfo  {journal} {Proceedings of the National Academy of Sciences}\ }\textbf {\bibinfo {volume} {119}},\ \bibinfo {pages} {e2119831119} (\bibinfo {year} {2022})}\BibitemShut {NoStop}%
\bibitem [{\citenamefont {Zhou}\ \emph {et~al.}(2023{\natexlab{b}})\citenamefont {Zhou}, \citenamefont {Zhdankin}, \citenamefont {Kunz}, \citenamefont {Loureiro},\ and\ \citenamefont {Uzdensky}}]{zhou2023magnetogenesis}%
  \BibitemOpen
  \bibfield  {author} {\bibinfo {author} {\bibfnamefont {M.}~\bibnamefont {Zhou}}, \bibinfo {author} {\bibfnamefont {V.}~\bibnamefont {Zhdankin}}, \bibinfo {author} {\bibfnamefont {M.~W.}\ \bibnamefont {Kunz}}, \bibinfo {author} {\bibfnamefont {N.~F.}\ \bibnamefont {Loureiro}},\ and\ \bibinfo {author} {\bibfnamefont {D.~A.}\ \bibnamefont {Uzdensky}},\ }\bibfield  {title} {\bibinfo {title} {Magnetogenesis in a collisionless plasma: from weibel instability to turbulent dynamo},\ }\href@noop {} {\bibfield  {journal} {\bibinfo  {journal} {The Astrophysical Journal}\ }\textbf {\bibinfo {volume} {960}},\ \bibinfo {pages} {12} (\bibinfo {year} {2023}{\natexlab{b}})}\BibitemShut {NoStop}%
\bibitem [{\citenamefont {Priest}\ and\ \citenamefont {Forbes}(2000)}]{priest_forbes_2000}%
  \BibitemOpen
  \bibfield  {author} {\bibinfo {author} {\bibfnamefont {E.}~\bibnamefont {Priest}}\ and\ \bibinfo {author} {\bibfnamefont {T.}~\bibnamefont {Forbes}},\ }\href {https://doi.org/10.1017/CBO9780511525087} {\emph {\bibinfo {title} {Magnetic Reconnection: MHD Theory and Applications}}}\ (\bibinfo  {publisher} {Cambridge University Press},\ \bibinfo {year} {2000})\BibitemShut {NoStop}%
\bibitem [{\citenamefont {Yamada}\ \emph {et~al.}(2010)\citenamefont {Yamada}, \citenamefont {Kulsrud},\ and\ \citenamefont {Ji}}]{yamada2010}%
  \BibitemOpen
  \bibfield  {author} {\bibinfo {author} {\bibfnamefont {M.}~\bibnamefont {Yamada}}, \bibinfo {author} {\bibfnamefont {R.}~\bibnamefont {Kulsrud}},\ and\ \bibinfo {author} {\bibfnamefont {H.}~\bibnamefont {Ji}},\ }\bibfield  {title} {\bibinfo {title} {Magnetic reconnection},\ }\href@noop {} {\bibfield  {journal} {\bibinfo  {journal} {Reviews of Modern Physics}\ }\textbf {\bibinfo {volume} {82}},\ \bibinfo {pages} {603} (\bibinfo {year} {2010})}\BibitemShut {NoStop}%
\bibitem [{\citenamefont {Hesse}\ and\ \citenamefont {Cassak}(2020)}]{review_2019}%
  \BibitemOpen
  \bibfield  {author} {\bibinfo {author} {\bibfnamefont {M.}~\bibnamefont {Hesse}}\ and\ \bibinfo {author} {\bibfnamefont {P.}~\bibnamefont {Cassak}},\ }\bibfield  {title} {\bibinfo {title} {Magnetic reconnection in the space sciences: Past, present, and future},\ }\href@noop {} {\bibfield  {journal} {\bibinfo  {journal} {Journal of Geophysical Research: Space Physics}\ }\textbf {\bibinfo {volume} {125}},\ \bibinfo {pages} {e2018JA025935} (\bibinfo {year} {2020})}\BibitemShut {NoStop}%
\bibitem [{\citenamefont {Kulsrud}\ and\ \citenamefont {Zweibel}(2008)}]{kulsrud2008origin}%
  \BibitemOpen
  \bibfield  {author} {\bibinfo {author} {\bibfnamefont {R.~M.}\ \bibnamefont {Kulsrud}}\ and\ \bibinfo {author} {\bibfnamefont {E.~G.}\ \bibnamefont {Zweibel}},\ }\bibfield  {title} {\bibinfo {title} {On the origin of cosmic magnetic fields},\ }\href@noop {} {\bibfield  {journal} {\bibinfo  {journal} {Reports on Progress in Physics}\ }\textbf {\bibinfo {volume} {71}},\ \bibinfo {pages} {046901} (\bibinfo {year} {2008})}\BibitemShut {NoStop}%
\bibitem [{\citenamefont {Weibel}(1959)}]{weibel1959spontaneously}%
  \BibitemOpen
  \bibfield  {author} {\bibinfo {author} {\bibfnamefont {E.~S.}\ \bibnamefont {Weibel}},\ }\bibfield  {title} {\bibinfo {title} {Spontaneously growing transverse waves in a plasma due to an anisotropic velocity distribution},\ }\href@noop {} {\bibfield  {journal} {\bibinfo  {journal} {Physical Review Letters}\ }\textbf {\bibinfo {volume} {2}},\ \bibinfo {pages} {83} (\bibinfo {year} {1959})}\BibitemShut {NoStop}%
\bibitem [{\citenamefont {Schep}\ \emph {et~al.}(1994)\citenamefont {Schep}, \citenamefont {Pegoraro},\ and\ \citenamefont {Kuvshinov}}]{schep1994generalized}%
  \BibitemOpen
  \bibfield  {author} {\bibinfo {author} {\bibfnamefont {T.~J.}\ \bibnamefont {Schep}}, \bibinfo {author} {\bibfnamefont {F.}~\bibnamefont {Pegoraro}},\ and\ \bibinfo {author} {\bibfnamefont {B.~N.}\ \bibnamefont {Kuvshinov}},\ }\bibfield  {title} {\bibinfo {title} {Generalized two-fluid theory of nonlinear magnetic structures},\ }\href@noop {} {\bibfield  {journal} {\bibinfo  {journal} {Physics of Plasmas}\ }\textbf {\bibinfo {volume} {1}},\ \bibinfo {pages} {2843} (\bibinfo {year} {1994})}\BibitemShut {NoStop}%
\bibitem [{\citenamefont {Wang}\ \emph {et~al.}(2015)\citenamefont {Wang}, \citenamefont {Kistler}, \citenamefont {Mouikis},\ and\ \citenamefont {Petrinec}}]{Wang2015}%
  \BibitemOpen
  \bibfield  {author} {\bibinfo {author} {\bibfnamefont {S.}~\bibnamefont {Wang}}, \bibinfo {author} {\bibfnamefont {L.~M.}\ \bibnamefont {Kistler}}, \bibinfo {author} {\bibfnamefont {C.~G.}\ \bibnamefont {Mouikis}},\ and\ \bibinfo {author} {\bibfnamefont {S.~M.}\ \bibnamefont {Petrinec}},\ }\bibfield  {title} {\bibinfo {title} {Dependence of the dayside magnetopause reconnection rate on local conditions},\ }\href {https://doi.org/https://doi.org/10.1002/2015JA021524} {\bibfield  {journal} {\bibinfo  {journal} {Journal of Geophysical Research: Space Physics}\ }\textbf {\bibinfo {volume} {120}},\ \bibinfo {pages} {6386} (\bibinfo {year} {2015})}\BibitemShut {NoStop}%
\bibitem [{\citenamefont {Zheng}\ and\ \citenamefont {Hu}(2018)}]{zheng2018observational}%
  \BibitemOpen
  \bibfield  {author} {\bibinfo {author} {\bibfnamefont {J.}~\bibnamefont {Zheng}}\ and\ \bibinfo {author} {\bibfnamefont {Q.}~\bibnamefont {Hu}},\ }\bibfield  {title} {\bibinfo {title} {Observational evidence for self-generation of small-scale magnetic flux ropes from intermittent solar wind turbulence},\ }\href@noop {} {\bibfield  {journal} {\bibinfo  {journal} {The Astrophysical Journal Letters}\ }\textbf {\bibinfo {volume} {852}},\ \bibinfo {pages} {L23} (\bibinfo {year} {2018})}\BibitemShut {NoStop}%
\bibitem [{\citenamefont {Chen}\ \emph {et~al.}(2019)\citenamefont {Chen}, \citenamefont {Hu},\ and\ \citenamefont {le~Roux}}]{chen2019analysis}%
  \BibitemOpen
  \bibfield  {author} {\bibinfo {author} {\bibfnamefont {Y.}~\bibnamefont {Chen}}, \bibinfo {author} {\bibfnamefont {Q.}~\bibnamefont {Hu}},\ and\ \bibinfo {author} {\bibfnamefont {J.~A.}\ \bibnamefont {le~Roux}},\ }\bibfield  {title} {\bibinfo {title} {Analysis of small-scale magnetic flux ropes covering the whole ulysses mission},\ }\href@noop {} {\bibfield  {journal} {\bibinfo  {journal} {The Astrophysical Journal}\ }\textbf {\bibinfo {volume} {881}},\ \bibinfo {pages} {58} (\bibinfo {year} {2019})}\BibitemShut {NoStop}%
\bibitem [{\citenamefont {Chen}\ and\ \citenamefont {Hu}(2022)}]{chen2022small}%
  \BibitemOpen
  \bibfield  {author} {\bibinfo {author} {\bibfnamefont {Y.}~\bibnamefont {Chen}}\ and\ \bibinfo {author} {\bibfnamefont {Q.}~\bibnamefont {Hu}},\ }\bibfield  {title} {\bibinfo {title} {Small-scale magnetic flux ropes and their properties based on in situ measurements from the parker solar probe},\ }\href@noop {} {\bibfield  {journal} {\bibinfo  {journal} {The Astrophysical Journal}\ }\textbf {\bibinfo {volume} {924}},\ \bibinfo {pages} {43} (\bibinfo {year} {2022})}\BibitemShut {NoStop}%
\bibitem [{\citenamefont {Drake}\ \emph {et~al.}(2021)\citenamefont {Drake}, \citenamefont {Agapitov}, \citenamefont {Swisdak}, \citenamefont {Badman}, \citenamefont {Bale}, \citenamefont {Horbury}, \citenamefont {Kasper}, \citenamefont {MacDowall}, \citenamefont {Mozer}, \citenamefont {Phan} \emph {et~al.}}]{drake2021switchbacks}%
  \BibitemOpen
  \bibfield  {author} {\bibinfo {author} {\bibfnamefont {J.}~\bibnamefont {Drake}}, \bibinfo {author} {\bibfnamefont {O.}~\bibnamefont {Agapitov}}, \bibinfo {author} {\bibfnamefont {M.}~\bibnamefont {Swisdak}}, \bibinfo {author} {\bibfnamefont {S.}~\bibnamefont {Badman}}, \bibinfo {author} {\bibfnamefont {S.}~\bibnamefont {Bale}}, \bibinfo {author} {\bibfnamefont {T.}~\bibnamefont {Horbury}}, \bibinfo {author} {\bibfnamefont {J.~C.}\ \bibnamefont {Kasper}}, \bibinfo {author} {\bibfnamefont {R.}~\bibnamefont {MacDowall}}, \bibinfo {author} {\bibfnamefont {F.}~\bibnamefont {Mozer}}, \bibinfo {author} {\bibfnamefont {T.}~\bibnamefont {Phan}}, \emph {et~al.},\ }\bibfield  {title} {\bibinfo {title} {Switchbacks as signatures of magnetic flux ropes generated by interchange reconnection in the corona},\ }\href@noop {} {\bibfield  {journal} {\bibinfo  {journal} {Astronomy \& Astrophysics}\ }\textbf {\bibinfo {volume} {650}},\ \bibinfo {pages} {A2} (\bibinfo {year} {2021})}\BibitemShut {NoStop}%
\bibitem [{\citenamefont {Zhou}\ \emph {et~al.}(2023{\natexlab{c}})\citenamefont {Zhou}, \citenamefont {Liu},\ and\ \citenamefont {Loureiro}}]{zhou2023electron}%
  \BibitemOpen
  \bibfield  {author} {\bibinfo {author} {\bibfnamefont {M.}~\bibnamefont {Zhou}}, \bibinfo {author} {\bibfnamefont {Z.}~\bibnamefont {Liu}},\ and\ \bibinfo {author} {\bibfnamefont {N.~F.}\ \bibnamefont {Loureiro}},\ }\bibfield  {title} {\bibinfo {title} {Electron heating in kinetic-{A}lfv{\'e}n-wave turbulence},\ }\href@noop {} {\bibfield  {journal} {\bibinfo  {journal} {Proceedings of the National Academy of Sciences}\ }\textbf {\bibinfo {volume} {120}},\ \bibinfo {pages} {e2220927120} (\bibinfo {year} {2023}{\natexlab{c}})}\BibitemShut {NoStop}%
\bibitem [{\citenamefont {Finn}\ and\ \citenamefont {Kaw}(1977)}]{finn1976coalescence}%
  \BibitemOpen
  \bibfield  {author} {\bibinfo {author} {\bibfnamefont {J.~M.}\ \bibnamefont {Finn}}\ and\ \bibinfo {author} {\bibfnamefont {P.~K.}\ \bibnamefont {Kaw}},\ }\bibfield  {title} {\bibinfo {title} {{Coalescence instability of magnetic islands}},\ }\href {https://doi.org/10.1063/1.861709} {\bibfield  {journal} {\bibinfo  {journal} {The Physics of Fluids}\ }\textbf {\bibinfo {volume} {20}},\ \bibinfo {pages} {72} (\bibinfo {year} {1977})}\BibitemShut {NoStop}%
\bibitem [{\citenamefont {Harris}(1962)}]{harris1962plasma}%
  \BibitemOpen
  \bibfield  {author} {\bibinfo {author} {\bibfnamefont {E.~G.}\ \bibnamefont {Harris}},\ }\bibfield  {title} {\bibinfo {title} {On a plasma sheath separating regions of oppositely directed magnetic field},\ }\href@noop {} {\bibfield  {journal} {\bibinfo  {journal} {Il Nuovo Cimento (1955-1965)}\ }\textbf {\bibinfo {volume} {23}},\ \bibinfo {pages} {115} (\bibinfo {year} {1962})}\BibitemShut {NoStop}%
\bibitem [{\citenamefont {Sweet}(1958)}]{sweet1958}%
  \BibitemOpen
  \bibfield  {author} {\bibinfo {author} {\bibfnamefont {P.}~\bibnamefont {Sweet}},\ }\bibfield  {title} {\bibinfo {title} {14. the neutral point theory of solar flares},\ }in\ \href@noop {} {\emph {\bibinfo {booktitle} {Symposium-International Astronomical Union}}},\ Vol.~\bibinfo {volume} {6}\ (\bibinfo {organization} {Cambridge University Press},\ \bibinfo {year} {1958})\ pp.\ \bibinfo {pages} {123--134}\BibitemShut {NoStop}%
\bibitem [{\citenamefont {Parker}(1957)}]{parker1957sweet}%
  \BibitemOpen
  \bibfield  {author} {\bibinfo {author} {\bibfnamefont {E.~N.}\ \bibnamefont {Parker}},\ }\bibfield  {title} {\bibinfo {title} {Sweet's mechanism for merging magnetic fields in conducting fluids},\ }\href@noop {} {\bibfield  {journal} {\bibinfo  {journal} {Journal of Geophysical Research}\ }\textbf {\bibinfo {volume} {62}},\ \bibinfo {pages} {509} (\bibinfo {year} {1957})}\BibitemShut {NoStop}%
\bibitem [{\citenamefont {Gro{\v{s}}elj}\ \emph {et~al.}(2017)\citenamefont {Gro{\v{s}}elj}, \citenamefont {Cerri}, \citenamefont {Navarro}, \citenamefont {Willmott}, \citenamefont {Told}, \citenamefont {Loureiro}, \citenamefont {Califano},\ and\ \citenamefont {Jenko}}]{grovselj2017fully}%
  \BibitemOpen
  \bibfield  {author} {\bibinfo {author} {\bibfnamefont {D.}~\bibnamefont {Gro{\v{s}}elj}}, \bibinfo {author} {\bibfnamefont {S.~S.}\ \bibnamefont {Cerri}}, \bibinfo {author} {\bibfnamefont {A.~B.}\ \bibnamefont {Navarro}}, \bibinfo {author} {\bibfnamefont {C.}~\bibnamefont {Willmott}}, \bibinfo {author} {\bibfnamefont {D.}~\bibnamefont {Told}}, \bibinfo {author} {\bibfnamefont {N.~F.}\ \bibnamefont {Loureiro}}, \bibinfo {author} {\bibfnamefont {F.}~\bibnamefont {Califano}},\ and\ \bibinfo {author} {\bibfnamefont {F.}~\bibnamefont {Jenko}},\ }\bibfield  {title} {\bibinfo {title} {Fully kinetic versus reduced-kinetic modeling of collisionless plasma turbulence},\ }\href@noop {} {\bibfield  {journal} {\bibinfo  {journal} {The Astrophysical Journal}\ }\textbf {\bibinfo {volume} {847}},\ \bibinfo {pages} {28} (\bibinfo {year} {2017})}\BibitemShut {NoStop}%
\bibitem [{\citenamefont {Fadeev}\ \emph {et~al.}(1965)\citenamefont {Fadeev}, \citenamefont {Kvabtskhava},\ and\ \citenamefont {Komarov}}]{fadeev1965self}%
  \BibitemOpen
  \bibfield  {author} {\bibinfo {author} {\bibfnamefont {V.}~\bibnamefont {Fadeev}}, \bibinfo {author} {\bibfnamefont {I.}~\bibnamefont {Kvabtskhava}},\ and\ \bibinfo {author} {\bibfnamefont {N.}~\bibnamefont {Komarov}},\ }\bibfield  {title} {\bibinfo {title} {Self-focusing of local plasma currents},\ }\href@noop {} {\bibfield  {journal} {\bibinfo  {journal} {Nuclear fusion}\ }\textbf {\bibinfo {volume} {5}},\ \bibinfo {pages} {202} (\bibinfo {year} {1965})}\BibitemShut {NoStop}%
\bibitem [{\citenamefont {Fonseca}\ \emph {et~al.}(2002)\citenamefont {Fonseca}, \citenamefont {Silva}, \citenamefont {Tsung}, \citenamefont {Decyk}, \citenamefont {Lu}, \citenamefont {Ren}, \citenamefont {Mori}, \citenamefont {Deng}, \citenamefont {Lee}, \citenamefont {Katsouleas} \emph {et~al.}}]{fonseca2002osiris}%
  \BibitemOpen
  \bibfield  {author} {\bibinfo {author} {\bibfnamefont {R.~A.}\ \bibnamefont {Fonseca}}, \bibinfo {author} {\bibfnamefont {L.~O.}\ \bibnamefont {Silva}}, \bibinfo {author} {\bibfnamefont {F.~S.}\ \bibnamefont {Tsung}}, \bibinfo {author} {\bibfnamefont {V.~K.}\ \bibnamefont {Decyk}}, \bibinfo {author} {\bibfnamefont {W.}~\bibnamefont {Lu}}, \bibinfo {author} {\bibfnamefont {C.}~\bibnamefont {Ren}}, \bibinfo {author} {\bibfnamefont {W.~B.}\ \bibnamefont {Mori}}, \bibinfo {author} {\bibfnamefont {S.}~\bibnamefont {Deng}}, \bibinfo {author} {\bibfnamefont {S.}~\bibnamefont {Lee}}, \bibinfo {author} {\bibfnamefont {T.}~\bibnamefont {Katsouleas}}, \emph {et~al.},\ }\bibfield  {title} {\bibinfo {title} {Osiris: A three-dimensional, fully relativistic particle in cell code for modeling plasma based accelerators},\ }in\ \href@noop {} {\emph {\bibinfo {booktitle} {Computational Science—ICCS 2002: International Conference Amsterdam, The Netherlands, April 21--24, 2002 Proceedings, Part III 2}}}\ (\bibinfo
  {organization} {Springer},\ \bibinfo {year} {2002})\ pp.\ \bibinfo {pages} {342--351}\BibitemShut {NoStop}%
\end{thebibliography}%

\clearpage

\foreach \n in {1,...,10} {
  \begin{figure}[htbp]
    \centering
    \includegraphics[width=\textwidth,page=1]{Supplemental_Material-pages-\n.pdf}
  \end{figure}
  \clearpage
}

\end{document}